\documentclass[11pt,reprint,showpacs,showkeys,nofootinbib,ssfamily]{article}
\usepackage{axodraw2}
\usepackage{amsmath,fontenc,calrsfs}
\usepackage{amsfonts,amssymb,mathtools,slashed,empheq}
\usepackage{cite} 
\usepackage{hyperref} 
\hypersetup{backref, colorlinks=true  } 
\usepackage{multirow}
\usepackage{float}
\usepackage{appendix}
\usepackage{color}
\usepackage{url}
\usepackage{subfigure}
\usepackage{footnote}
\usepackage{authblk} 
\newcommand*{\email}[1]{\normalsize{#1}\par} 

\def\beq{\begin{equation}}
\def\eeq{\end{equation}}
\def\bea{\begin{eqnarray}}
\def\eea{\end{eqnarray}}
\def\nn{\nonumber}

\def\chic1{\chi_{c1}}

\newcommand{\ep}{\epsilon}
\newcommand{\ga}{\alpha}
\newcommand{\gs}{\sigma}

\newcommand{\ts}{\tilde{\sigma}}
\newcommand{\ta}{\tilde{\alpha}}
\newcommand{\ve}{\varepsilon}

\newcommand{\la}{\langle}
\newcommand{\ra}{\rangle}

\def \Im{\text{Im}\,}

\def\Xint#1{\mathchoice
   {\XXint\displaystyle\textstyle{#1}}%
   {\XXint\textstyle\scriptstyle{#1}}%
   {\XXint\scriptstyle\scriptscriptstyle{#1}}%
   {\XXint\scriptscriptstyle\scriptscriptstyle{#1}}%
   \!\int}
\def\XXint#1#2#3{{\setbox0=\hbox{$#1{#2#3}{\int}$}
    \vcenter{\hbox{$#2#3$}}\kern-.5\wd0}}

\def\dashint{\Xint-}

\def\Xint#1{\mathchoice
   {\XXint\displaystyle\textstyle{#1}}%
   {\XXint\textstyle\scriptstyle{#1}}%
   {\XXint\scriptstyle\scriptscriptstyle{#1}}%
   {\XXint\scriptscriptstyle\scriptscriptstyle{#1}}%
   \!\int}

\newcommand{\va}{\mathbf{a}}

\newcommand{\hvk}{\hat{\mathbf{k}}}
\newcommand{\vk}{\mathbf{k}}
\newcommand{\hvp}{\hat{\mathbf{p}}}
\newcommand{\vp}{\mathbf{p}}
\newcommand{\hvq}{\hat{\mathbf{q}}}
\newcommand{\vq}{\mathbf{q}}

\newcommand{\hvz}{\hat{\mathbf{z}}}
\newcommand{\vz}{\mathbf{z}}

\newcommand{\cZ}{\cal{Z}}

\newcommand{\cL}{\cal{L}}

\newcommand{\cE}{\cal{E}}
\newcommand{\bcE}{{\bar{\cal{E}}}}

\newcommand{\cH}{\cal{H}}

\usepackage{hyperref}
\hypersetup{%
  colorlinks = true,
  linkcolor  = black
}

\title{Nuclear matter from the ladder resummation in terms of the experimental nucleon-nucleon scattering amplitudes}

\author[]{J.~M.~Alarc\'on}
\affil[]{\small  Universidad de Alcal\'a, Grupo de F\'{\i}sica Nuclear y de Part\'{\i}culas, \\
\small   Departamento de F\'{\i}sica y Matem\'aticas,  28805 Alcal\'a de Henares (Madrid), Spain\\
\email{\small jmanuel.alarcon@uah.es}}
\author[]{J.~A. Oller}
\affil[]{\small  Departamento de F\'{\i}sica, Universidad de Murcia, 30071 Murcia, Spain \\ \email{\small oller@um.es}}

\begin{document}
\maketitle
\begin{abstract}
  Infinite nuclear matter is studied by resuming the series of ladder diagrams based on the results developed by us in Ann.~Phys.~437,~168741~(2022). The master formula for the energy density is explicitly solved for the case of contact interactions, within a pionless description of the nucleon-nucleon interactions. Renormalized results are obtained which are directly expressed in terms of the nucleon-nucleon phase shifts and mixing angles in partial-wave amplitudes up to an including $G$ waves, with convergence reached under the inclusion of higher partial waves. The energy per particle, density and sound velocity resulting from the ladder series are given for symmetric and neutron matter. 
  This resummation of the ladder diagrams provides a rigorous result that may be used as low-density reference for other parameterizations of $\bcE$ for higher densities.

\end{abstract}



\section{Introduction}

The raise of precision physics has brought a need for more rigorous calculations in the low energy sector of QCD. In order to interpret such experiments' outcome correctly and claim a possible discovery, theoretical calculations with controlled systematic errors are crucial. The properties of baryonic matter have been studied for such kind of programs. In that case, there is additional complication besides dealing with low energy QCD, since the interactions occur in a baryonic environment, and vacuum approaches are not applicable.

In this paper we  apply many-body field theory to the calculation of the energy per particle $\bcE$
for nuclear matter, and other magnitudes that can be deduced thereof.
We define nuclear matter as  an infinite uniform system of nucleons
interacting by the strong force without electromagnetic interactions. This system is 
supposed to approximate  the interior of a heavy nucleus.
The proportion of protons and neutrons in the system is controlled by the fraction of
protons $x_p$, so that for $x_p=0$ one has pure neutron matter (PNM) and $x_p=1/2$ corresponds to symmetric nuclear
matter (SNM), being both extremes of special interest in our research here.
Indeed, the equation of state of nuclear matter is nowadays one of the most active fields where these types of calculations are necessary, especially in the study of neutron stars and gravitational waves.  The former offers a unique possibility of studying nuclear matter under extreme conditions and test our current theoretical approaches for baryonic matter.

The many-body calculations within  perturbation theory \cite{fetter} 
are well-known   
since long \cite{Huang:1957im,Lee:1957zza,efimov:1965,efimov:1968,Baker:1971vm,bishop:1973,Hammer:2000xg}. 
However, for larger scattering lengths
the perturbative expansion in powers of $a_0 k_F$ fails, with $a_0$ the $S$-wave scattering length.
Of course, this is the case if one is interested in the unitary limit $|a_0 k_F|\to\infty$ \cite{zwerger.200901.1,Giorgini2008TheoryOU,Randeria2013BCSBECCA}, which  is closely related to neutron matter due to the large and negative neutron-neutron ($nn$)  scattering length   $a_{nn}=-18.95\pm 0.40$~fm \cite{Chen:2008zzj}. Note that   $|a_{nn}|\gg m_\pi^{-1}$, with $m_\pi$ the pion mass and which inverse typically  controls the longest range of strong interactions. 

A time-honored  possibility to end with a  meaningful result for large scattering lengths  is to resum the two-body interactions in the medium \cite{fetter}.  
 In the Brueckner theory \cite{bethebrueckner,Brueckner:1954zz,Brueckner:1955nst,Brueckner:1955zze,Hu:2016nkw}  the infinite series of interacting particle-particle intermediate states  is resummed, where the two particles always have momenta above their Fermi momenta.\footnote{The Fermi momenta are globally denoted  by $k_F$ or $\xi$.} This theory was generalized by Thouless \cite{Thouless:1960anp} considering also two-fermion intermediate states with momenta below the Fermi momenta (or intermediate hole-hole states). The notation of ladder diagrams was also introduced by him to denote the associated Feynman graphs. As a result, both particle-particle and hole-hole intermediate states interact between two consecutive rungs of the ladder series, and their infinite iteration is resummed. 
 The ladder resummation at zero temperature  is studied Refs.~\cite{Steele:2000qt,Lacour:2009ej,Kaiser:2011cg,Kaiser:2012sr},  taking into account Pauli blocking without including self-energy effects. Resumming the ladder diagrams in such circumstances is typically considered a good starting point for calculating $\bcE$ \cite{Steele:2000qt,Schafer:2005kg,Drischler:2017wtt}, also supported by the power counting arguments of Ref.~\cite{Lacour:2009ej,Oller:2009zt,Meissner:2001gz}.

When the interactions between two spin-1/2 fermions is reduced to its scattering length, an algebraic renormalized formula for the ladder resummation was accomplished by Kaiser in Ref.~\cite{Kaiser:2011cg}.
 Within only dimensional regularization, the extension of the previous result for taking also into account the contributions from the  effective range in $S$ wave was obtained by the same author in Ref.~\cite{Kaiser:2012sr} where, due to off-shell effects, the resulting formula  was conjectured and checked up to some order. The case of an interaction given by the $P$-wave scattering volume $a_1$ was separately discussed in the same reference, and the resummation within dimensional regularization was accomplished.  
 The connection between the ladder resummation and the density-functional theory  in many-body calculations has been studied in  \cite{Boulet:2019wfd,Grasso:2018pen}. Furthermore, it is conjectured  \cite{Kolck2017UnitarityAD,Konig:2016utl} that for nuclear and atomic systems with two-body interactions near the unitary limit, the binding energy of the three-body system defines the relevant scale for low-energy observables, such as particle energy.

  We recently resummed the ladder diagrams for arbitrary spin-1/2 fermion-fermion interactions in vacuum in Ref.~\cite{Alarcon:2021kpx}. The resummation can take into account higher orders in the effective range expansion (ERE) of a partial-wave amplitude (PWA) and/or any number of PWAs. 
  The case of contact interactions is fully resolved and  renormalized results for $\bar{\cE}$ are obtained, so that they are directly expressed in terms of vacuum scattering parameters of the ERE.  
  In the present work we proceed further, and derive the needed equations for different Fermi momenta $\xi_1$ (protons) and $\xi_2$ (neutrons). In addition, we give the expression for $\bcE$ when infinitely many orders are included in the ERE for (un)coupled PWAs, such that the resulting $\bcE$ is directly expressed in terms of phase shifts and mixing angles.  In this way, the results from the ladder resummation are completely independent of cutoff and have no free parameters. We have then explored the cases of SNM and PNM, discussing $\bcE$ for both cases, and its first and second order derivatives, namely, the pressure (or the equation of state) and the sound velocity. For the case of PNM our results at low densities have been extrapolated towards larger densities by using a quadratic expression in $x_p$. Results compatible with nowadays constraints and determinations are obtained for the symmetry energy $S_0$ and its logarithmic slope in density $L$ at nuclear matter saturation.

  The contents of the manuscript are organized as follows. After this Introduction, the resummation of the ladder diagrams and its partial-wave decomposition are discussed in Sec.~\ref{sec.200904.1}. An important needed element is the in-medium nucleon-nucleon scattering amplitude which is discussed in Sec.~\ref{sec.190808.1}, and solved in Sec.~\ref{sec.200406.1} for the case of contact interactions. The results for SNM and PNM are given in Sec.~\ref{sec.221013.1}. The last section contains a summary and  concluding remarks.


\section{Resummation of ladder diagrams for the energy density $\cE$}
\label{sec.200904.1}

\def\theequation{\arabic{section}.\arabic{equation}}
\setcounter{equation}{0}

The resummation of the ladder diagrams for evaluating $\cE$ in terms of an arbitrary fermion-fermion vacuum $T$-matrix was accomplished by us in Ref.~\cite{Alarcon:2021kpx}.
This derivation was based on the many-body formalism of Ref.~\cite{Oller:2001sn},
which we refer as the in-medium many-body quantum field theory.
Since the resummation of ladder diagrams was derived in detail in Ref.~\cite{Alarcon:2021kpx} here we only provide a brief summary signaling the main steps in the derivation. We also briefly recap the power counting of Ref.~\cite{Oller:2009zt} for in-medium calculations. 

\subsection{Summary of the in-medium many-body formalism of Ref.~\cite{Oller:2001sn}}
\label{sec.201223.1}

Reference~\cite{Oller:2001sn} determines the in-medium Lagrangian after integrating out the fermions in the nuclear medium.
This is accomplished by calculating the generating functional ${\cZ}[J]$ of in-medium Green functions 
with external sources $J$'s.

The vacuum Lagrangian  contains a pure bosonic part, ${\cal L}_\phi$, and another 
bilinear in the fermion fields, that is globally called ${\cal L}_{\bar{\psi}\psi}=\bar{\psi}D\psi$.
The operator $D$, which we write as $D=D_0-A$, comprises the free fermion Lagrangian $D_0=i\gamma^\mu\partial_\mu-m$, with $m$ the nucleon mass in the isospin limit, and the interacting part $A$, which incorporates the
boson-fermion interactions and external sources.  The bosons can be either light, e.g. pions,
or heavy ones which, when integrated out, give rise to contact multi-fermion interactions.
In this way, we do not need to additionally incorporate monomials with extra fermion fields in the Lagrangian density,
like quartic ones  ${\cal L}_{\bar{\psi}\bar{\psi}\psi\psi}$ and so on. 

The result for $ e^{i{\cZ}[J]}$ calculated in Ref.~\cite{Oller:2001sn} can be written as
\begin{align}
\label{190521.7}
&e^{i{\cZ}[J]}=\!\int\! [dU] \exp\Big[i\!\int\! dx\, {\cL}_{\phi}
  -i\!\int\!\! \frac{d\vp}{(2\pi)^3 } 
\!\int\! {\rm Tr}\Big( A[I-D_0^{-1}A]^{-1}\arrowvert_{(x,y)}
n(p)\Big) dx\,dy\, e^{ip(x-y)} \\ 
&-\frac{1}{2}(-i)^2\!\int \frac{d\vp}{(2\pi)^3}\!\int\!  \frac{d\vq}{(2\pi)^3}
\!\int\! {\rm Tr}\Big(
A[I-D_0^{-1}A]^{-1}\arrowvert_{(x,x')}n(q)A[I-  D_0^{-1}A]^{-1}\arrowvert_{(y',y)}n(p)\Big)\nn\\
&\times
 e^{ip(x-y)} e^{-iq(x'-y')} dx\,dx'\,dy\,dy' +...\Big ]~,\nn
\end{align}
such that the exponent in the integrand is $i$ times the total in-medium Lagrangian.
In this equation each trace is taken over the spin and other internal indices of the fermions,
like the isospin ones, and  $D_0^{-1}$ is the free vacuum fermion propagator,  
\begin{align}
\label{200904.1}
iD_0^{-1}(p)=\frac{i}{p^0-E(p)+i\ep}~.
\end{align} 
The functions $n(p)$ restrict the momentum $\vp$ below the Fermi momentum $\xi_\alpha$ for each nucleon species $\alpha$, with $\alpha=1(2)$ referring to a proton(neutron).\footnote{The Fermi momentum could also depend on the nucleon but 
  we do not consider further this case because our interest here rests in unpolarized Fermi systems.} By employing a matrix notation in the isospin space we can write
\begin{align}
\label{190521.4b}
n(p)&=\left(
\begin{matrix}
\theta({\xi}_{1}-|\vp|) & 0\\
0 & \theta({\xi}_{2}-|\vp|)
\end{matrix}
\right)~,
\end{align}
 where  $\theta(x)$ is the Heaviside or step function.

 Each term in the sum  in Eq.~\eqref{190521.7} involving at least one $n(p)$ is denoted as an in-medium generalized vertex (IGV), after Ref.~\cite{Oller:2001sn}, and its total number is called $V_\rho$.
 The IGVs are made by sewing  non-local vacuum vertices  $\Gamma$,
\begin{align}
\label{190521.8}
\Gamma & \equiv-iA[I-D_0^{-1}A]^{-1}=-iA\sum_{n=0}^\infty (D_0^{-1}A)^n~,
\end{align}
with Fermi seas, with each of them involving a factor $n(p)$ and a sum over all the states in the Fermi seas (which implies an integration over momentum).
In addition, there is a numerical factor  $(-1)^{n+1}/n$ from the series of $\ln(1+\ve)$, 
\begin{align}
\label{190521.4}
\ln(1+\ve)&=\ve-\frac{\ve^2}{2}+\frac{\ve^3}{3}-\frac{\ve^4}{4}+\ldots=\sum_{n=1}^\infty (-1)^{n+1}\frac{\ve^n}{n}~.
\end{align}

Equation.~\eqref{190521.7} gives rise to Feynman rules and graphs.
The associated propagators for the fermion lines are either in-medium insertions of on-shell Fermi seas,
connecting $\Gamma$ vertices, or fermion vacuum propagators joining
vacuum vertices $A$.  
In the following, a pure vacuum fermion propagator  $iD_0^{-1}(p)$ is depicted as a solid line, and
a Fermi-sea insertion $n(p)(2\pi)\delta(p^0-\vp^2/2m)$ is drawn by a double line.
In both cases one has to sum over spin and isospin indices, 
and integrate over the intermediate four-momentum $\int d^4p/(2\pi)^4$.
Each vertex $-iA$ 
is plotted as a filled circle, while the non-local $\Gamma$ vertices are  plotted as empty circles.
Additionally, one should keep in mind that bosonic and source lines can stem from the $A$ vertices.
 Of course,  we refer to the original Ref.~\cite{Oller:2001sn} for the derivation and more extensive discussion of this many-body framework. A good illustration is the pure perturbative calculations done in Ref.~\cite{Meissner:2001gz}, see also Refs.~\cite{Goda:2013bka,Goda:2013npa}, and nonperturbative ones were undertaken in Refs.~\cite{Oller:2009zt,Lacour:2010ci,Lacour:2009ej,Dobado:2011gd}. For a recent review see Ref.~\cite{Oller:2019ssq}.

\begin{figure}
  \begin{center}
    \begin{tabular}{lccccr}
 \includegraphics[width=.215\textwidth]{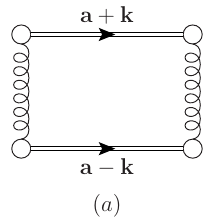}  &&&&& 
  \includegraphics[width=.5\textwidth]{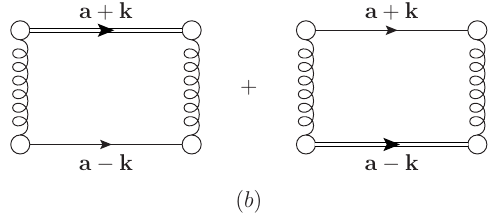} 
  \end{tabular}
\end{center}
\caption{{\small In  panel (a) we show $L_d(p,\va)$ and in panel (b)  $L_m(p,\va)$, which comprises two Feynman diagrams. The double lines correspond to Fermi seas insertions, made up of on-shell fermions with momentum below their respective Fermi momentum. Here, a spring schematically indicates any expanded interaction with momentum flow $\pm \vk$ along it.}
\label{fig.220408.1} 
}
\end{figure} 

\subsection{Fock and Hartree diagram contributions to ${\cal E}$}
\label{sec.190603.1}

Reference~\cite{Alarcon:2021kpx} performs the resummation of the ladder diagrams to calculate the energy density $\cE$ of a system of fermions of spin 1/2 with an arbitrary vacuum fermion-fermion $T$ matrix, which we call $t$.
Here, we first give the formula obtained in Ref.~\cite{Alarcon:2021kpx}, and then introduce the different operators that appear in it.

The resulting expression for the interacting part of $\cE$ in the ladder approximation, $\cE_{\cL}$, is \cite{Alarcon:2021kpx}
\begin{align}
\label{190615.1b}
\cE_{\cL}&=i{\rm Tr}\left(
\sum_{n=1}^{\infty}\frac{(t_mL_d)^{n}}{2n}\right)=-\frac{i}{2}{\rm Tr}\ln\left(I-t_m L_d\right)\, ,
\end{align}
with the series fixing the branch of $\ln z$, with ${\rm arg}z\in (-\pi,\pi)$. In this equation the in-medium fermion-fermion $T$ matrix is denoted by $t_m$, while $L_d$ is a unitary loop function made up of two Fermi-sea insertions, which is shown in the panel (a) of Fig.~\ref{fig.220408.1}.

In addition to the interacting part one also has  to sum the densities of kinetic energies, $\cE_K$, of protons and neutrons  
\begin{align}
\label{220407.3}
  \cE_K&=\frac{\xi_1^5}{10m\pi^2}+\frac{\xi_2^5}{10m\pi^2}=\rho_1 \frac{3\xi_1^2}{10m}+\rho_2\frac{3\xi_2^2}{10m}~.
\end{align}

Given the four-momenta $k_1$ and $k_2$ of the two fermions we introduce the four vectors 
\begin{align}
\label{190615.2}
a&=\frac{1}{2}(k_1+k_2)~,\\
k&=\frac{1}{2}(k_1-k_2)~,\nn
\end{align}
so that 
\begin{align}
  \label{220624.1}
  k_1=a+k~,~k_2=a-k
\end{align}
For the on-shell case  we use $\vp$ instead of $\vk$ to denote the relative momentum, with $p\equiv |\vp|$, so that $k_i^0=E(\vp_i)=p_i^2/2m$. 
It is important to keep in mind that the total four-momentum $a$ is conserved during the in-medium scattering process of two fermions because of translational symmetry. 

There are two important in-medium unitary functions. One is $L_d(p,\va)$, already mentioned, and the other is $L_m(p,\va)$, which consists of two mixed intermediate states composed by a Fermi sea insertion and a vacuum propagator.
The loop function $L_m(p,\va)$  is depicted in the panel (b) of  Fig.~\ref{fig.220408.1}. These loop-function operators are given by the expressions,
\begin{align}
\label{190615.7a}
L_d(p,\va)&=\frac{i}{2}\sum_{\sigma,\alpha}\int\frac{d^3k_1}{(2\pi)^3}\frac{d^3k_2}{(2\pi)^3}
\theta(\xi_{\alpha_1}-|\vk_1|)\theta(\xi_{\alpha_2}-|\vk_2|)(2\pi)^4\delta(k_1+k_2-2a)\nn\\
&\times |\vk_1\sigma_1\alpha_1,\vk_2\sigma_2\alpha_2{\rangle_A}\, {_A\langle} \vk_1 \sigma_1\alpha_1,\vk_2\sigma_2\alpha_2|~,\\
\label{190615.3a}
L_m(p,\va)&=\frac{1}{2}\sum_{\sigma,\alpha}\int\frac{d^3k}{(2\pi)^3}
\frac{\theta(\xi_{\alpha_1}-|\va+\vk|)+\theta(\xi_{\alpha_2}-|\va-\vk|)}{2a^0-\frac{|\va+\vk|^2}{2m}
  -\frac{|\va-\vk|^2}{2m}+i\ep}\nn\\
&\times |\vk_1\sigma_1\alpha_1,\vk_2\sigma_2\alpha_2{\rangle_A}\, {_A\langle} \vk_1 \sigma_1\alpha_1,\vk_2\sigma_2\alpha_2|~.
\end{align}
Here, we have denoted by $|\vk_1\sigma_1\alpha_1,\vk_2\sigma_2\alpha_2{\rangle_A}$ the antisymmetric two-fermion intermediate state with momenta $\vk_1=\va+\vk$, $\vk_2=\va-\vk$, third components of spin $\sigma_1$, $\sigma_2$, and third components of isospin $\alpha_1$, $\alpha_2$. The sum over the spin and isospin indices $\sigma_i$ and $\alpha_i$ is denoted by $\displaystyle{\sum_{\sigma,\alpha}}$. A  symmetry factor 1/2 is included in Eqs.~\eqref{190615.7a} and \eqref{190615.3a} because the two-fermion state is antisymmetric.

Since $\va$ is conserved we  express in the following the two-fermion intermediate states simply in terms of its relative momentum $\vk$, as $|\vk\sigma_1\sigma_2\alpha_1\alpha_2{\rangle_A}$.
Further, as the in-medium states summed over in the trace of Eq.~\eqref{190615.1b} are on-shell, it follows that 
\begin{align}
\label{190615.4}
2ma^0-\va^2&=m\big(E(\va+\vp)+E(\va-\vp)\big)-\va^2=\vp^2  ~.
\end{align}
Then, we can simplify the expressions for $L_d(p,\va)$ and $L_m(p,\va)$ as 
\begin{align}
\label{190615.7}
L_d(p,\va)&=i\frac{m p}{16\pi^2}\sum_{\sigma,\alpha}\int d\hat{\vk}\,\theta(\xi_{\alpha_1}-|\va+p \hat{\vk}|)
\theta(\xi_{\alpha_2}+|\va-p\hat{\vk}|)\,
|p\hat{\vk} \sigma_1\sigma_2\alpha_1\alpha_2{\rangle_A}\,{_A\langle}p\hat{\vk} \sigma_1\sigma_2\alpha_1\alpha_2|~,
\\
\label{190615.3}
L_m(p,\va)&=- \frac{m}{2}\sum_{\sigma,\alpha}\int\frac{d^3k}{(2\pi)^3}
\frac{ \theta(\xi_{\alpha_1}-|\va+\vk|) + \theta(\xi_{\alpha_2}-|\va-\vk|) }{ \vk^2-\vp^2-i\ep }
|\vk\sigma_1\sigma_2\alpha_1\alpha_2{\rangle_A}\,{_A\langle}\vk \sigma_1\sigma_2\alpha_1\alpha_2|~.
\end{align}

In terms of the vacuum $T$-matrix $t$ and $L_m$, the operational equation that defines $t_m(\va)$ is \cite{Alarcon:2021kpx}
\begin{align}
\label{190624.2}
t_m(\va)&=t+t L_m(p,\va) t_m(\va)~.
\end{align}
In this way, the in-medium $T$ matrix $t_m(\va)$ results by iterating $t$
with mixed intermediate states making up $L_m$. 
From this equation  the matrix elements of $t_m(\va)$ between the  initial and final two-fermion antisymmetric states,
$|\vp\sigma_1\sigma_2\alpha_1\alpha_2\ra $ and $|\vp'\sigma'_1\sigma'_2\alpha'_1\alpha'_2\ra$,  respectively, fulfill the integral equation (IE) 
\begin{align}
\label{190624.3b}
&\la \vp'\sigma'_1\sigma'_2\alpha'_1\alpha'_2|t_m(\va)|\vp\sigma_1\sigma_2\alpha_1\alpha_2\ra
=\la \vp'\sigma'_1\sigma'_2\alpha'_1\alpha'_2|t(\va)|\vp\sigma_1\sigma_2\alpha_1\alpha_2\ra
\\
&-\frac{m}{2}\sum_{\ts,\ta} \int \frac{d^3 k}{(2\pi)^3}
\la \vp'\sigma'_1\sigma'_2\alpha'_1\alpha'_2|t(\va)|\vk\ts_1\ts_2\ta_1\ta_2\ra
\frac{\theta(\xi_1-|\va+\vk|)+\theta(\xi_2-|\va-\vk|)}{\vk^2-\vp^2-i\ep}\nn\\
&\times \la \vk\ts_1\ts_2\ta_1\ta_2|t_m(\va)|\vp\sigma_1\sigma_2\alpha_1\alpha_2\ra~,\nn
\end{align}
where we have used the expression for $L_m(p,\va)$ in Eq.~\eqref{190615.3}.

Reference~\cite{Alarcon:2021kpx} demonstrates in Sec.~2.3.2 that, despite the complex nature of the operators $t_m$ and $L_d$  and the explicit presence of the imaginary unity in  Eq.~\eqref{190615.1b},  $\cE_{\cL}$  is real for the case of equal Fermi momenta. The demonstration is rather technical and we omit it here for brevity and to avoid repeating ourselves with Ref.~\cite{Alarcon:2021kpx}. The basic point is that the argument of the $\ln$ in Eq.~\eqref{190615.1b} can be diagonalized and its eigenvalues are phase factors of unite modulus. This is why the $\arctan$ series found in Refs.~\cite{Kaiser:2011cg,Kaiser:2012sr} always appear in these calculations.

\subsection{Power counting}
\label{sec.220409.1}

Reference~\cite{Oller:2009zt} develops a low-energy power counting for nuclear matter, with the fermion-fermion interactions  counted as ${\cal O}(1)$.
Low-energy nucleon-nucleon interactions fall into this category because the scattering lengths are unnaturally large, and pion exchange also is counted as O(1) since it is proportional to the linear momentum exchanged squared times the pion propagator. In this process a fermion energy $~\vp^2/2m$ is counted as ${\cal O}(k_F^2)$, and then a fermion propagator as ${\cal O}(k_F^{-2})$, which also applies to a Fermi-sea insertion within an IGV. 
 In presenting this power counting one has to distinguish between exchanges of light bosons (referred as $\pi$), responsible for the long-range parts of the fermionic interactions, and heavier bosons (referred as $H$), which give rise to short-range interactions.
Actually for the latter we have in mind  the limit of infinite mass, so that at the end one has contact interactions.
Related to this, one does not either need to consider their presence in ${\cal L}_\phi$, which is restricted to the light fields.
In the vertices of type $A$ from ${\cal L}_{\bar{\psi}\psi}$ and those in ${\cal L}_\phi$ we count the number of derivatives and of bosonic lines attach to each of them. 
In this way, we indicate by $\nu_i$ the number of bosons (heavy and light)  attached to the $i_{\rm th}$ bilinear vertex, of which $\omega_i$ are heavy fields, and by $d_i$ its number of derivatives.
Concerning the purely bosonic vertices  from ${\cal L}_\phi$ we denote by $n_i$ the number of light fields in $i_{\rm th}$ vertex, and by $\delta_i$ the number of derivatives there.
Finally, the total number of vertices from ${\cal L}_{\bar{\psi}\psi}$ and ${\cal L}_\phi$ is called $V$ and $V_\pi$, respectively, and the total number of external light bosonic lines is called $E_\pi$.

With this preamble one can calculate straightforwardly the chiral dimension  $\nu$ of an in-medium diagram, i.e. the power to which the typical size of the momentum involved in the diagram is raised.
The original derivation can be found in Ref.~\cite{Oller:2009zt}, being reviewed and simplified in Ref.~\cite{Oller:2019ssq}. The resulting expression for the power counting is

\begin{align}
  \label{220409.1}
  \nu&=3-E_\pi+\sum_{i=1}^{V_\pi}(\delta_i+n_i-4)+\sum_{i=1}^N(d_i+\nu_i+\omega_i-2)+V_\rho~,
\end{align}

For the calculation of the interacting part of $\cE$ at least two fermions are involved in the interaction, so that $V_\rho\geq 2$,
and there are no external light fields, and hence $E_\pi=0$. The leading-order (LO) contribution, involving two-body fermionic interactions, requires $V_\rho=2$, which corresponds to the integration over the two Fermi seas in the calculation of the trace in Eq.~\eqref{190615.1b} for computing $\cE_{\cL}$.
In this way,  according to the counting of Eq.~\eqref{220409.1}, $\cE$  has dominant chiral order $\nu=5$.
This is shown in Ref.~\cite{Alarcon:2021kpx} with explicit algebraic expressions, e.g. when evaluating the $S$-wave contribution to ${\cE}_{\cL}$, and it will also be clear in our applications discussed in Secs.~\ref{sec.220403.1} and \ref{sec.220403.2}.
Higher order contributions arise by increasing $V_\rho$ or any of the coefficients inside parentheses concerning the number of bosonic lines or derivatives, so that the combinations in parenthesis  become positive, instead of being zero as  for the LO contributions.


Let us also notice that for small values of $k_F$, such that it is much smaller than the light-field masses, we can also consider the latter as heavy fields and run into the limit of only contact interactions. This is a limit of special significance for the applications developed below, and also for nuclear physics in general. Then, the power counting in Eq.~\eqref{220409.1} simplifies to
\begin{align}
  \label{220410.1}
\nu&=3+\sum_{i=1}^N(d_i+2\omega_i-2)+V_\rho~.
\end{align}
Notice that the parenthesis is $\geq 0$ as long as $d_i\geq 0$ because $\omega_i\geq 1$. The LO contributions to $\cE$ are those with $V_\rho=2$ and vanishing combinations inside parenthesis.  In the applications discussed in  Secs.~\ref{sec.220403.2} and \ref{sec.220403.1}  we actually go beyond the LO contributions because the  vacuum nucleon-nucleon interactions are given in terms of their phenomenological phase shifts and mixing angles, despite in-medium corrections are implemented at LO by resumming the ladder diagrams. 

A posteriori, by attending to the change of the results of $\bcE$ for the case of SNM under the variation of the  the Gaussian cutoff, cf. Fig.~\ref{fig.220823.1}, we find a value for the scale $\Lambda$ around 350~MeV. This number stems from having an uncertainty of 1.5~MeV for a value of $\bcE$  around $-3.5$~MeV at $k_F=150$~MeV. Then, according to the power counting in Eq.~\eqref{220410.1}, the NLO in-medium correction that requires $V_\rho=3$  is suppressed by an extra power of $k_F/\Lambda$, from where $\Lambda\approx 350$~MeV results. The variation in the results with the Gaussian cutoff of $\bcE$ for PNM are smaller, cf. Fig.~\ref{fig.220824.pnm},  and the resulting $\Lambda$ is larger.

\subsection{Partial-wave expansion}
\label{sec.190630.1}
\def\theequation{\arabic{section}.\arabic{equation}}

This subsection corresponds basically to Sec.~3 of Ref.~\cite{Alarcon:2021kpx}, to which we refer for more details. The only addition here, which indeed is rather straightforward to implement, consists of taking into account the isospin degrees of freedom. 
 Within the notation developed so far  we can rewrite the Eq.~\eqref{190615.1b} as
\begin{align}
\label{190630.2}
&{\cE_{\cH}}=-i\int \frac{pdp}{m\pi}\int \frac{d^3a}{\pi^3}
{\rm Tr}\left(\ln\left[I-t_m(\va) L_d(p,\va)\right]\right)\\
\label{190630.3}
&=-\frac{i}{2}\sum_{C}\int\!\! \frac{pdp}{m\pi}\int \frac{d^3a}{\pi^3}
\int \frac{d^3 q}{(2\pi)^3}
{_A\langle} \vq,C|\ln\left[I-t_m(\va) L_d(p,\va)\right]{|\vq,C\rangle_A}~,
\end{align}
and $C$ has the same meaning as set of spin and isospin indices as
in Eq.~\eqref{190624.3b}. The extra factor of 1/2 in the last equation is
introduced due to the antisymmetrized nature of the two-fermion states, indicated by the
subscript $A$ in the bra and kets.

Equation~\eqref{190630.3} can be simplified by noting the fact that the matrix elements of $L_d(p,\va)$ are diagonal in
the absolute value of the three-momentum, as it is clear from Eq.~\eqref{190615.7}. 
 Explicitly, the matrix elements of $L_d(p,\va)$ are 
\begin{align}
\label{190709.3}
\langle \vq'\beta'_1\beta'_2|L_d(p,\va)|\vq\beta_1\beta_2\rangle&=
(2\pi^2)^2\frac{\delta(q'-p)\delta(q-p)}{p^4}\langle \hvq'\beta'_1\beta'_2 |\tilde{L}_d(p,\va)|\hvq\beta_1\beta_2\rangle~, \\
\langle \hvq'\beta'_1\beta'_2 |\tilde{L}_d(p,\va)|\hvq\beta_1\beta_2\rangle
&=2imp \delta(\hvq'-\hvq)\delta_{\beta'_1\beta_1}\delta_{\beta'_2\beta_2}\theta(\xi_1-|\va+p\hvq|)\theta(\xi_2-|\va-p\hvq|)~.
\nn
\end{align}
 The factor  $2\pi^2\delta(q'-q)/p^2$, corresponding to the identity operator in the $p$ space, can be factored out and Eq.~\eqref{190630.3} for $\cE_{\cL}$ can be rewritten as
\begin{align}
{\cE_{\cL}}&=-\frac{i}{2}\sum_{\sigma_{1,2},\alpha_{1,2}}\int\!\! \frac{pdp}{m\pi}\int \frac{d^3a}{\pi^3}\int\frac{d\hvp}{4\pi}
\,{_A\langle} \vp,\sigma_1\sigma_2\alpha_1\alpha_2|\ln\left[I-t_m(\va) {L}_d(p,\va)\right]|\vp,\sigma_1\sigma_2\alpha_1\alpha_2 {\rangle_A}~,
\end{align}
where the tilde on top of $L_d$ is dropped to ease the notation.

Another simplification in the expression for $\cE_{\cL}$ comes from rotational symmetry, so that one can take always $\va$ along the $\hvz$ axis. In this way, the angular integration over $\va$ is just a factor $4\pi$, and the following simplified expression results \cite{Alarcon:2021kpx}
\begin{align}
  \label{220404.1}
{\cE}_{\cL}&=-2i\sum_{\sigma_{1,2},\alpha_{1,2}}\int\!\! \frac{pdp}{m\pi}\int \frac{a^2da}{\pi^2}\int\frac{d\hvp}{4\pi}
\,{_A\langle} \vp,\sigma_1\sigma_2\alpha_1\alpha_2|\ln\left[I-t_m(a\hvz) {L}_d(p,a\hvz)\right]|\vp,\sigma_1\sigma_2\alpha_1\alpha_2 {\rangle_A}~.
  \end{align}

Next, we make a PWA expansion in the relative-motion variables $\vp$ in terms of the partial-wave vector states 
$|J\mu\ell SI i_3,p,\ra$, where $J$ is the total angular momentum, $\mu$ is its third component, $\ell$ is the orbital angular momentum, $S$ is the total spin, $I$ is the total isospin, and $i_3$ is its third component.
One has to take into account the value for
the scalar product between a partial-wave vector $|J\mu\ell SI i_3p\ra$
and the plane-wave ones $|\vp,\sigma_1\sigma_2\alpha_1\alpha_2\ra_A$.
 The relation between both bases is 
\begin{align}
  \label{200501.3}
&|\vp, \sigma_1\sigma_2\alpha_1\alpha_2\rangle_{\!A}\!\!=\!\!\sqrt{4 \pi}\sum_{J\mu\ell S}
( \sigma_1 \sigma_2 \sigma_3| s_1 s_2 S)
(m \sigma_3 \mu | \ell S J) (\alpha_1\alpha_2 i_3|\tau_1\tau_2 I) {Y_\ell^m}(\hvp)^* \chi( S \ell I) |J \mu \ell S I i_3 p \rangle,\\
&{_A\langle} \vp',\sigma_1\sigma_2\alpha_1\alpha_2|J\mu\ell S I i_3 p\rangle=\chi(S\ell)
\frac{4\pi^\frac{5}{2}\delta(p'-p)}{p^2}(\sigma_1\sigma_2 s_3|s_1s_2s)(ms_3 \mu|\ell S J)
(\alpha_1\alpha_2 i_3|\tau_1\tau_2 I)
Y_\ell^m(\hvp)~,\nn\\
&\chi(S\ell I)=\frac{1-(-1)^{\ell+S+I}}{\sqrt{2}}~.\nn
\end{align}
The factor $\chi(S\ell I)$ is non-zero for odd $\ell+S+I$, as required by Fermi statistics for a two-fermion state.  
When inserted the PWA expansion in Eq.~\eqref{220404.1} we just focus on the angular variables since
the expression is already diagonal in $p$. After taking into account the standard orthogonality properties for the Clebsch-Gordan coefficients and spherical harmonics \cite{Alarcon:2021nwd} we then have that
\begin{align}
{\cE_{\cL}}
\label{200501.6b}
&=-\frac{2i}{m\pi^3}\!\!\!\!\sum_{{\scriptsize \begin{array}{l}J,\mu,\ell\\ S,I,i_3\end{array}}}\!\!\!\!\chi(S\ell I)^2
\int_0^\infty pdp\int_0^\infty a^2da
\la J\mu\ell S I i_3 p|\ln\left[I-t_m(a\hvz) L_d(p,a\hvz)\right]
|J\mu\ell S I i_3 p\ra~.
\end{align}
Let us remark that the presence of $L_d(p,a\hvz)$ limits the possible values of $a$ and $p$ since it requires that  $\xi_1^2+\xi_2^2\geq 2(a^2+p^2)\geq 0$.


\section{Integral equations for $t_m(a\hvz)$}
\label{sec.190808.1}
\def\theequation{\arabic{section}.\arabic{equation}}
\setcounter{equation}{0}

Equation~\eqref{190624.2} for $t_m(\va)$ can be formally solved as
\begin{align}
\label{190625.1}
t_m(\va)=\left[I-t L_m(p,\va)\right]^{-1}t~,
\end{align}
 and $t_m(\va)^{-1}$ then fulfills 
\begin{align}
\label{190625.2}
t_m(\va)^{-1}&=t^{-1}-L_m(p,\va)~.
\end{align}
In turn,  the vacuum $T$-matrix $t$ satisfies a Lippmann-Schwinger  equation
\begin{align}
\label{190625.3}
t^{-1}&=V^{-1}+G(p)~.
\end{align}
Here, $V$ is the potential and $G(p)$ is the vacuum unitarity loop function with intermediate states involving two fermions, 
\begin{align}
\label{190625.3b}
 G(p)&=-\frac{m}{2}\int\frac{d^3k}{(2\pi)^3}\frac{|\vk{\rangle_A}\,{_A\langle} \vk|}{k^2-p^2-i\ep}~.
\end{align}
Therefore, from Eqs.~\eqref{190625.2} and \eqref{190625.3} we can also express $t_m(\va)$ as
\begin{align}
\label{190625.3c}
  t_m(\va)^{-1}&=V^{-1}+G(p)-L_m(p,\va)~.
\end{align}
  
Given the formal solution  for $t_m(\va)$ in Eq.~\eqref{190625.3c}, it is clear that
this in-medium $T$ matrix satisfies also the operational equation  
\begin{align}
  \label{190808.1}
t_m(\va)&=V-V[G(p)-L_m(p,\va)]t_m(\va)~,
\end{align}
for a given  $E=p^2/m$. 

In order to arrive to the IE for $t_m(\va)$ let us first analyze the matrix elements of the
kernel $G(p)-L_m(p,\va)$ between antisymmetrized plane-wave states.
We take the expressions in Eqs.~\eqref{190625.3b} and \eqref{190615.3} for the operators
$G(p)$ and $L_m(p,\va)$, respectively.
From the operator $G(p)$ we then have from Eq.~\eqref{190625.3b} that
\begin{align}
\label{200925.3}
{_A\la}\vk' \ts'_1 \ts'_2 \ta'_1 \ta'_2 | G(p) | \vk \ts_1\ts_2\ta_1\ta_2 {\ra_A}
=
-\frac{m(2\pi)^3}{k^2-p^2-i\ep}
&\left[
  \delta(\vk'-\vk) \delta_{\ts'_1\ts_1}\delta_{\ts'_2\ts_2}\delta_{\ta'_1\ta_1}\delta_{\ta'_2\ta_2}\right.\\
  &\left.-
\delta(\vk'+\vk)\delta_{\ts'_1\ts_2}\delta_{\ts'_2\ts_1}\delta_{\ta'_1\ta_2}\delta_{\ta'_2\ta_1})\right]~.\nn
\end{align}
For the case of the operator $L_m(p,\va)$, Eq.~\eqref{190615.3}, more care is needed because the dependence of the
Fermi momenta on the isospin indices of the intermediate states, $\ta_1$ and $\ta_2$. Namely,
\begin{align}
\label{200925.4}
&{_A\la}\vk' \ts'_1 \ts'_2 \ta'_1 \ta'_2 |L_m(p,\va)| \vk \ts_1\ts_2\ta_1\ta_2 {\ra_A}=-
\frac{(2\pi)^3m}{k^2-p^2-i\ep}
\left[\delta(\vk'-\vk) \delta_{\ts'_1\ts_1}\delta_{\ts'_2\ts_2}\delta_{\ta'_1\ta_1}\delta_{\ta'_2\ta_2}\right.\\
  &\left.-
  \delta(\vk'+\vk)\delta_{\ts'_1\ts_2}\delta_{\ts'_2\ts_1}\delta_{\ta'_1\ta_2}\delta_{\ta'_2\ta_1})\right]\nn
 \left[\theta(\xi_{\ta_1}-|\vk+\va|)+\theta(\xi_{\ta_2}-|\vk-\va|)\right]~.
\end{align}
The sum over the two Heaviside functions factorizes because
the sum over the intermediate states is symmetric under the simultaneous
exchange of the subscripts $1\leftrightarrow 2$ and $\vq\to -\vq$. 

Putting together these results
we end with the following expression for the operator $-V[G(p)-L_m(p,\va)]t_m(\va)$,
\begin{align}
\label{200926.1}
&-V[G(p)-L_m(p,\va)]t_m(\va)=\\
&-\frac{1}{2^2}\!\!\sum_{\ts,\ts',\ta,\ta'}\int\!\!\frac{d^3k'}{(2\pi)^3}\frac{d^3k}{(2\pi)^3}V|\vk'\ts'_1\ts'_2\ta'_1\ta'_2{\ra_A}
{_A\la}\vk'\ts'_1\ts'_2\ta'_1\ta'_2|G(p)\!-\!L_m(p,\va)|\vk,\ts_1\ts_2\ta_1\ta_2{\ra_A}
{_A\la}\vk\ts_1\ts_2\ta_1\ta_2|t_m(\va)\nn\\
&=\frac{m}{2}\sum_{\ts,\ta}\int\frac{d^3k}{(2\pi)^3}V|\vk\ts_1\ts_2\ta_1\ta_2{\ra_A}
\frac{1-\theta(\xi_{\ta_1}-|\vk+\va|)-\theta(\xi_{\ta_2}-|\vk-\va|)}{k^2-p^2-i\ep}
{_A\la}\vk\ts_1\ts_2\ta_1\ta_2|t_m(\va)~.\nn
\end{align}
  Then, taking the previous result in  Eq.~\eqref{190808.1}, 
the IE for the two-body scattering operator in momentum space reads
\begin{align}
\label{200925.5}
&{_A\la}\vp' \gs'_1 \gs'_2 \ga'_1 \ga'_2 |t_m(\va)| \vp \gs_1\gs_2\ga_1\ga_2 {\ra_A}
={_A\la}\vp' \gs'_1 \gs'_2 \ga'_1 \ga'_2 |V| \vp \gs_1\gs_2\ga_1\ga_2 {\ra_A}\\
&+\frac{m}{2}\sum_{\ts,\ta}\int\frac{d^3k}{(2\pi)^3}
{_A\la}\vp' \gs'_1 \gs'_2 \ga'_1 \ga'_2 |V| \vk \ts_1\ts_2\ta_1\ta_2 {\ra_A}
\frac{1-\theta(\xi_{\ta_1}-|\vk+\va|)-\theta(\xi_{\ta_2}-|\vk-\va|)}{k^2-p^2-i\ep}\nn \\
&\times {_A\la}\vk \ts_1 \ts_2 \ta_1 \ta_2 |t_m(\va)| \vp \gs_1\gs_2\ga_1\ga_2 {\ra_A}~. \nn 
\end{align}


From the decomposition of the antisymmetrized plane-wave states in the partial-wave basis,
Eq.~\eqref{200501.3},  the Eq.~\eqref{200926.1} for $-V[G(p)-L_m(p,\va)]t_m(\va)$  in the partial-wave basis is
\begin{align}
\label{200926.2}
&-V[G(p)-L_m(p,\va)]t_m(\va)=\\
&m\!\sum_{\ts,\ta}\!\!\! \sum_{{\scriptsize \begin{array}{l} J \mu \ell m_3\\ S s_3 I i_3 \end{array}} }\!\!\!\!
\sum_{{\scriptsize \begin{array}{l}J'\mu'\ell'm'_3 \\S's'_3I'i'_3\end{array}}}\!\!\!
\int\frac{d^3k}{(2\pi)^2}V|J'\mu'\ell'S'I'i'_3 k \ra
\frac{1-\theta(\xi_{\ta_1}-|\vk+\va|)-\theta(\xi_{\ta_2}-|\vk-\va|)}{k^2-p^2-i\ep}\la J\mu\ell S Ii_3 k |t_m(\va)\nn\\
&\times \chi(S\ell I)\chi(S'\ell' I')(\ts_1\ts_2s'_3|s s S')(\ts_1\ts_2s_3|s s S)
(\ta_1\ta_2i'_3|\tau\tau I')(\ta_1\ta_2i_3|\tau\tau I)(m'_3s'_3\mu'|\ell' S' J')(m_3s_3\mu|\ell S J)\nn\\
&\times Y_{\ell'}^{m'_3}(\hvk)^*Y_{\ell}^{m_3}(\hvk)~,\nn
\end{align}
where $s=\tau=1/2$ is the spin and isospin of the each nucleon, respectively. It is also clear from
this equation that $i_3=i_3'=\ta_1+\ta_2$
The sum over $\ts_1$ and $\ts_2$ can be readily done because of 
 the orthogonality properties of the Clebsch-Gordan coefficients, 
\begin{align}
\label{200926.3}
\sum_{\ts_1,\ts_2}(\ts_1\ts_2s'_3|s s S')(\ts_1\ts_2s_3|s s S)=\delta_{s'_3s_3}\delta_{S'S}~.
\end{align}

In the following we choose $\displaystyle{\va}$ along the $z$ axis because this is enough to calculate $\cE_{\cL}$,
cf. Eq.~\eqref{200501.6b}, and it
also induces extra simplifications in the final IE for $t_m(a\hvz)$.
Because of this choice it is clear that there is no dependence on the azimuthal angle of $\vk$ in the
integral of Eq.~\eqref{200926.1}, because $|\vk \pm a\hvz|$ only depends on its polar angle.
Thus,
\begin{align}
\label{200926.4}
\int_0^{2\pi} d\varphi Y_{\ell'}^{m'_3}(\theta,\varphi)^*  Y_{\ell}^{m_3}(\theta,\varphi)\propto
\delta_{m'_3 m_3}~.
\end{align}
As a result $\mu'=\mu$ because $\mu'=s'_3+m'_3=s_3+m_3=\mu$. 
 Then, we can get rid of the sums over
$\ts_1$, $\ts_2$, $S'$, $s'_3$, $m'_3$, $\mu'$ and $i'_3$ in Eq.~\eqref{200926.2}, which then becomes 
\begin{align}
\label{200926.4}
&-V[G(p)-L_m(p,\va)]t_m(a\vz)= \nn\\
&m\!\sum_{\ta_1,\ta_2}\!\!\!\sum_{{\scriptsize \begin{array}{l}J\mu\ell m_3 \\ Ss_3 J'\ell'\end{array}}}
\!\!\!\sum_{{\scriptsize \begin{array}{l} I'Ii_3\end{array}}}
\int\frac{d^3k}{(2\pi)^2}V|J'\mu\ell'SI'i_3 k \ra
\frac{1-\theta(\xi_{\ta_1}-|\vk+a\hvz|)-\theta(\xi_{\ta_2}-|\vk-a\hvz|)}{k^2-p^2-i\ep}
\la J\mu\ell S Ii_3 k |t_m(a\hvz)\nn\\ 
&\times \chi(S\ell I)\chi(S\ell' I')(\ta_1\ta_2i_3|\tau\tau I')(\ta_1\ta_2i_3|\tau\tau I)
(m_3s_3\mu|\ell' S J')(m_3s_3\mu|\ell S J)Y_{\ell'}^{m_3}(\hvk)^*Y_{\ell}^{m_3}(\hvk)~.
\end{align}

\subsection{Integral equation for $t_m(a\hvz)$ in partial waves: General values for $\xi_1,$ $\xi_2$}
\label{sec.201114.1}

Now, we deduce the integral equation for $t_m(a\hvz)$ for general values of $\xi_1$ and $\xi_2$, so that $\xi_1$ and $\xi_2$ are not assumed to be equal as it was the case in Ref.~\cite{Alarcon:2021kpx}. This generalization is only relevant for $i_3=0$, since for $i_3=\pm 1$ then $\ta_1=\ta_2=\pm 1/2$, respectively, and for these cases one can take directly result from Sec.~4 of Ref.~\cite{Alarcon:2021kpx}.

We then continue with the case $i_3=0$ and deduce the corresponding IE for the $t_m(a\hvz)$, and take first 
Eq.~\eqref{200926.4} with $I'\neq I$, $i_3=0$. For $\tau=1/2$ we then have the following substructure within the integrand,
\begin{align}
\label{201114.5}
&\frac{1-(-1)^{I'-I}}{2} \sum_{\ta_1,\ta_2}(\ta_1\ta_2i_3|\tau\tau I')(\ta_1\ta_2i_3|\tau\tau I)
\left[1-\theta(\xi_{\ta_1}-|\vk+a\hvz|)-\theta(\xi_{\ta_2}-|\vk-a\hvz|)\right]\\
&=\frac{1-(-1)^{I'-I}}{4}\left[-\theta(\xi_{1}-|\vk+a\hvz|)-\theta(\xi_{2}-|\vk-a\hvz|)
+\theta(\xi_{2}-|\vk+a\hvz|)+\theta(\xi_{1}-|\vk-a\hvz|)
  \right]~,\nn
\end{align}
where  in order to simplify the notation  the Fermi momenta $\xi_{\ta_i}$ are denoted as
\begin{align}
\label{201114.8}
\xi_1\equiv \xi_{+\frac{1}{2}}~,\\
\xi_2\equiv \xi_{-\frac{1}{2}}~.\nn
\end{align}

Next, we notice that $(-1)^{\ell'}=-(-1)^\ell$ because $I'\neq I$, the total spin $S$ is the same and there is contribution only when the Fermi statistics factors $\chi(S\ell I), \chi(S\ell' I')\neq 0$.
Then, by exchanging $\vk\to -\vk$ in the last two step functions in Eq.~\eqref{201114.5}, taking into account the parity rule for the spherical harmonics, $Y_\ell^m(-\vk)=(-1)^\ell Y_\ell^m(\vk)$,  the contributions in Eq.~\eqref{200926.4} with $I'\neq I$ become
\begin{align}
\label{201114.7}
&-m\!\!\!\!\sum_{{\scriptsize \begin{array}{l}J\mu\ell m_3 \\ Ss_3Ii_3\end{array}}}
\!\!\!\sum_{{\scriptsize \begin{array}{l}J'\ell' I'\end{array}}}
\frac{1-(-1)^{I'-I}}{2}\int\frac{d^3k}{(2\pi)^2}
\frac{V|J'\mu\ell'SI'i_3 k \ra \la J\mu\ell S Ii_3 k |t_m(a\hvz)}{k^2-p^2-i\ep}
Y_{\ell'}^{m_3}(\hvk)^*Y_{\ell}^{m_3}(\hvk)\\
&\times \chi(S\ell I)\chi(S\ell' I')
(m_3s_3\mu|\ell' S J')(m_3s_3\mu|\ell S J) \left[\theta(\xi_{1}-|\vk+a\hvz|)+\theta(\xi_{2}-|\vk-a\hvz|)\right]\nn ~.
\end{align}
On the other hand, for those with $I'=I$ we have  that $(\ta_1\ta_2 i_3|\tau\tau I)^2=1/2$ in
all cases, and the sum over the isospin indices $\ta_1$ and $\ta_2$ gives
\begin{align}
\label{201114.9}
&\frac{1+(-1)^{I-I'}}{4}\chi(S\ell I)\chi(S \ell' I')
\left[2-\theta(\xi_1-|\vk+a\hvz|)-\theta(\xi_2-|\vk-a\hvz|)
-\theta(\xi_2-|\vk+a\hvz|)\right.\\
&\left.-\theta(\xi_1-|\vk-a\hvz|)\right]Y_{\ell'}^{m_3}(\vk)^*Y_{\ell}^{m_3}(\vk)~.\nn
\end{align}
Since now $(-1)^{\ell}=(-1)^{\ell'}$ because $S$ is conserved, $I'=I$ and there is contribution only for $\chi(S\ell I),\chi(S\ell'I')\neq 0$,  the exchange $\vk\to-\vk$ implies that  the contributions in Eq.~\eqref{200926.4} with $I'=I$ read
\begin{align}
\label{201114.11}
&m\!\!\!\sum_{{\scriptsize \begin{array}{l}J\mu\ell m_3 \\ Ss_3Ii_3\end{array}}}
\!\!\!\!\sum_{{\scriptsize \begin{array}{l}J'\ell' I' \end{array}}}
\!\!\frac{1+(-1)^{I'-I}}{2} \int\frac{d^3k}{(2\pi)^2} \chi(S\ell I)\chi(S\ell' I')
(m_3s_3\mu|\ell' S J')(m_3s_3\mu|\ell S J)Y_{\ell'}^{m_3}(\hvk)^*Y_{\ell}^{m_3}(\hvk)\nn \\
&\times V|J'\mu\ell'SI'i_3 k\ra\frac{1-\theta(\xi_1-|\vk+a\hvz|)-\theta(\xi_2-|\vk-a\hvz|)}{k^2-p^2-i\ep}
\la J\mu\ell S Ii_3 k |t_m(a\hvz)~.
\end{align}

Putting together Eqs.~\eqref{201114.7} and \eqref{201114.11}, the resulting IE  reads
\begin{align}
\label{201114.12b}
&\la J'\mu\ell'SI'i_3 p'|t_m(a\hvz)|J\mu\ell S I i_3 p\ra=
\la J'\mu\ell'SI'i_3 p'|V|J\mu\ell S I i_3 p\ra
+m\!\!\!\sum_{{\scriptsize \begin{array}{l}J_1\ell_1 m_3 \\ s_3 \ell_2 I_1 \end{array}}}
\!\!\int\frac{d^3k}{(2\pi)^2}\\
&\times
\frac{ \chi(S\ell_2 I')\chi(S\ell_1 I_1)}{k^2-p^2-i\ep}
(m_3s_3\mu|\ell_2 S J')(m_3s_3\mu|\ell_1 S J_1)Y_{\ell_2}^{m_3}(\hvk)^*Y_{\ell_1}^{m_3}(\hvk) \la J'\mu\ell'SI'i_3 p'|V|J'\mu\ell_2SI'i_3 k\ra
\nn \\
&\times 
\la J_1\mu\ell_1 S I_1i_3 k |t_m(a\hvz)|J\mu\ell S I i_3 p\ra
\left\{\frac{1+(-1)^{I'-I_1}}{2}-\theta(\xi_1-|\vk+a\hvz|)-\theta(\xi_2-|\vk-a\hvz|)\right\}
~,\nn
\end{align}
where we have taken into account that the potential $V$ conserves isospin.
Now, since the combination $1+(-1)^{I'-I_1}$ conserves isospin, this implies that
\begin{align}
  \label{220813.1}
  \int d\vk Y_{\ell_2}^{m_3}(\hvk)^* Y_{\ell_1}^{m_3}(\hvk)\chi(S\ell_2I')
  \chi(S\ell_1I_1)\frac{1+(-1)^{I'-I_1}}{2}=  \int d\vk Y_{\ell_2}^{m_3}(\hvk)^* Y_{\ell_1}^{m_3}(\hvk)\chi(S\ell_2I')
  \chi(S\ell_1I_1)~,
\end{align}
by simply exchanging $\hvk\to-\hvk$ in the original integral, and taking into account that Fermi statistics requires then that $(-1)^{\ell_2}=(-1)^{\ell_1}$ for $I'=I_1$, and $(-1)^{\ell_2}=-(-1)^{\ell_1}$ for $I'\neq I_1$.  
 Then, we can simplify Eq.~\eqref{201114.12b} as
\begin{align}
\label{201114.12}
&\la J'\mu\ell'SI'i_3 p'|t_m(a\hvz)|J\mu\ell S I i_3 p\ra=
\la J'\mu\ell'SI'i_3 p'|V|J\mu\ell S I i_3 p\ra
+m\!\!\!\sum_{{\scriptsize \begin{array}{l}J_1\ell_1 m_3 \\ s_3 \ell_2 I_1 \end{array}}}
\!\!\int\frac{d^3k}{(2\pi)^2}\\
&\times
\frac{ \chi(S\ell_2 I')\chi(S\ell_1 I_1)}{k^2-p^2-i\ep}
(m_3s_3\mu|\ell_2 S J')(m_3s_3\mu|\ell_1 S J_1)Y_{\ell_2}^{m_3}(\hvk)^*Y_{\ell_1}^{m_3}(\hvk) \la J'\mu\ell'SI'i_3 p'|V|J'\mu\ell_2SI'i_3 k\ra
\nn \\
&\times 
\la J_1\mu\ell_1 S I_1i_3 k |t_m(a\hvz)|J\mu\ell S I i_3 p\ra
\left\{1-\theta(\xi_1-|\vk+a\hvz|)-\theta(\xi_2-|\vk-a\hvz|)\right\}
~.\nn
\end{align}

We can write the IE in Eq.~\eqref{220813.1} in a more compact matrix form for   as
\begin{align}
\label{201114.13}
&[t_m(a\hvz)](p',p)=[V](p',p)
+\frac{m}{(2\pi)^2}\int_0^\infty\frac{k^2 dk}{k^2-p^2-i\ep}
[V](p',k)\, {\cal A} \,[t_m(a\hvz)](k,p)~,
\end{align}
with the matrices
\begin{align}
\label{201114.14}
&[V](p',k)_{J'\ell'I',J_2\ell_2I_2}
=\delta_{J' J_2}\delta_{I' I_2}\langle J'\mu \ell' S I' i_3 p'|V|J_2\mu\ell_2 S I_2 i_3 k\rangle~,
\\
\label{201114.15}
&[t_m(a\hvz)](k,p)_{J_1\ell_1 I_1,J\ell I}=
\langle J_1\mu\ell_1 S I_1 i_3 k|t_m(a\hvz)|J\mu\ell S I i_3 p\rangle~,
\\
\label{201114.16}
&{\cal A}_{J_2\mu\ell_2 I_2, J_1\mu\ell_1 I_1}=\chi(S\ell_2 I_2)\chi(S\ell_1I_1)\sum_{m_3 s_3}
(m_3s_3\mu|\ell_2SJ_2)(m_3s_3\mu|\ell_1SJ_1) \int d\hvk Y_{\ell_2}^{m_3}(\hvk)^*
Y_{\ell_1}^{m_3}(\hvk)\\
&\times \left\{1-\theta(\xi_1-|\vk+a\hvz|)-\theta(\xi_2-|\vk-a\hvz|)
\right\}~.\nn
\end{align}

We also notice that our final expressions for the IE obeyed by $t_m(a\hvz)$,  Eqs.~\eqref{201114.12} and \eqref{201114.16}, are also applicable  when  $i_3=\pm 1$ by just replacing
\begin{align}
  \label{220813.2}
\theta(\xi_1-|\vk+a\hvz|)+\theta(\xi_2-|\vk-a\hvz|)\to 2\theta(\xi_{\frac{i_3}{2}}-|\vk-a\hvz|)~,  
\end{align}
which is the case studied in Ref.~\cite{Alarcon:2021kpx}.

\subsubsection{Some symmetry properties of PWAs: General values for $\xi_1$, $\xi_2$}
\label{sec.201115.1}

 Let us show that we do not really need to calculate the PWAs of $t_m(a\hvz)$ with negative
$\mu$ since they obey the rule 
\begin{align}
\label{201115.1}
\langle J'-\mu\ell' SIi_3p'|t_m(a\hvz)|J-\mu\ell SIi_3p\rangle
&=(-1)^{\ell'+\ell+J'+J}\langle J'\mu\ell' SI'i_3 p'|t_m(a\hvz)|J\mu\ell S I i_3 p\rangle
\end{align}
To prove it  we start by considering the IE of Eq.~\eqref{201114.12} for the PWAs with $-\mu$.
Because of rotational symmetry the matrix elements of $V$ are independent of $\mu$.
We also employ the symmetry property of the Clebsch-Gordan coefficients under the change of sign of the third components of spin \cite{rose.190706.1}, then 
\begin{align}
\label{201115.2}
(m_3\sigma_3-\mu|\ell_2SJ')(m_3\sigma_3-\mu|\ell_1SJ_1)&=(-1)^{\ell_2+\ell_1+J'+J_1}
(-m_3-\sigma_3 \mu|\ell_2SJ')(-m_3-\sigma_3 \mu|\ell_1SJ_1)~.
\end{align}
Next, we take into account that the product of the two spherical harmonics is real and we can write
\begin{align}
\label{201115.3}
Y_{\ell_2}^{m_3}(\hvk)^*Y_{\ell_1}^{m_3}(\vk)=Y_{\ell_2}^{m_3}(\hvk)Y_{\ell_1}^{m_3}(\vk)^*=Y_{\ell_2}^{-m_3}(\hvk)^*Y_{\ell_1}^{-m_3}(\vk)~,
\end{align}
where we have also exchanged the sign of the third components of the angular momenta by making use of the well-known property $Y_\ell^m(\hvk)^*=(-1)^mY_\ell^{-m}(\hvk)$.

Implementing this procedure, the IE for $(-1)^{\ell'+\ell+J'+J}\la J'-\mu\ell' SIi_3p'|t_m(a\hvz)|J-\mu\ell SIi_3p \ra$ from Eq.~\eqref{201114.12} becomes
\begin{align}
\label{201115.4}
&(-1)^{\ell'+\ell+J'+J}\la J'-\mu\ell'SI'i_3 p'|t_m(a\hvz)|J-\mu\ell S I i_3 p\ra=
\underbrace{(-1)^{\ell'+\ell+J'+J}}_{+1} \underbrace{\la J'\mu \ell'SI'i_3 p'|V|J\mu\ell S I i_3 p\ra}_{\propto \delta_{J'J}\delta_{\ell',\ell+\rm{mod}(2)}}\nn\\
&+m\!\!\!\sum_{{\scriptsize \begin{array}{l}J_1\ell_1 m_3 \\ s_3 \ell_2 I_1 \end{array}}}
\!\!\int\frac{d^3k}{(2\pi)^2}
\frac{ \chi(S\ell_2 I')\chi(S\ell_1 I_1)}{k^2-p^2-i\ep}
(-m_3-s_3\mu|\ell_2 S J')(-m_3-s_3\mu|\ell_1 S J_1)Y_{\ell_2}^{-m_3}(\hvk)^*Y_{\ell_1}^{-m_3}(\hvk)
\nn \\
&\times \underbrace{(-1)^{J'+\ell'+J'+\ell_2}}_{+1}\underbrace{\la J'\mu\ell'SI'i_3 p'|V|J'\mu\ell_2SI'i_3 k\ra}_{\propto \delta_{\ell',\ell_2+\rm{mod}(2)}}
(-1)^{J_1+\ell_1+J+\ell}\la J_1\mu\ell_1 S I_1i_3 k |t_m(a\hvz)|J\mu\ell S I i_3 p\ra\nn\\
&\times
\left[1-\theta(\xi_1-|\vk+a\hvz|)-\theta(\xi_2-|\vk-a\hvz|)\right]
~.
\end{align}
This IE, after relabeling the dummy indices $-m_3\to m_3$ and $-s_3\to s_3$ (which also has a symmetric sum interval around zero), is actually the same IE as the one satisfied by $\la J'\mu\ell' SI'i_3 p'|t_m(a\hvz)|J\mu\ell S I i_3 p\ra$, and Eq.~\eqref{201115.1} follows.

As a corollary of Eq.~\eqref{201115.1} we notice that for $\mu_1=0$ then
it is necessary that  
\begin{align}
\label{201115.5}
(-1)^{J'+J}=(-1)^{\ell'+\ell}
\end{align}
otherwise the PWA is zero.

 The PWAs of $t_m(a\hvz)$ are symmetric under the exchange of the
initial and final quantum numbers, namely, 
\begin{align}
\label{201115.6}
\langle J'\mu\ell' SI'i_3p|t_m(a\hvz)|J\mu\ell SIi_3p'\rangle
&=\langle J\mu\ell SIi_3p'|t_m(a\hvz)|J'\mu\ell' SI'i_3p\rangle
\end{align}
with the scattering energy fixed by $p$, so that $E=p^2/m$. 
This relation is particularly useful for the on-shell case with $p'=p$, which is the one
needed in the evaluation of ${\cal E}_{\cH}$. It implies then that the in-medium on-shell $T$ matrix is symmetric
under the exchange of the discrete labels.

For the  demonstration we use that the matrix elements of $V$ because of time reversal are invariant under the exchange of the initial and final states between them. Therefore, the IE for $\langle J\mu\ell SIi_3p|t_m(a\hvz)|J'\mu\ell' SI'i_3p'\rangle$ from Eq.~\eqref{201114.12} reads
\begin{align}
\label{201115.7}
&\la J\mu\ell SI i_3 p|t_m(a\hvz)|J'\mu\ell' S I' i_3 p'\ra=
\la J'\mu\ell'SI'i_3 p'|V|J\mu\ell S I i_3 p\ra
+m\!\!\!\!\!\!\sum_{{\scriptsize \begin{array}{l}J_1\ell_1 m_3 \\ s_3 \ell_2 I_1 \end{array}}}
\!\!\!\!\int\frac{d^3k}{(2\pi)^2}\frac{ \chi(S\ell_2 I')\chi(S\ell_1 I_1)}{k^2-p^2-i\ep}\nn\\
&\times
(m_3s_3\mu|\ell_2 S J)(m_3s_3\mu|\ell_1 S J_1)\underbrace{Y_{\ell_2}^{m_3}(\hvk)Y_{\ell_1}^{m_3}(\hvk)^*}_{\begin{array}{l}\text{\tiny It is real. The complex conjugate is taken}\end{array}} \la J_1\mu\ell_1 S I_1i_3 k |t_m(a\hvz)|J'\mu\ell' S I' i_3 p'\ra
\nn \\
&\times 
\la J \mu\ell_2SI i_3 k|V|J\mu\ell SI i_3 p\ra 
\left[1-\theta(\xi_1-|\vk+a\hvz|)-\theta(\xi_2-|\vk-a\hvz|)\right]
~.
\end{align}
This is the same IE as the one satisfied by
$\langle J'\mu\ell'SI'i_3p'|t_m(a\hvz)|J\mu \ell SIi_3p\rangle$ as we wanted to show.
In order to arrive to this conclusion we have used
the fact that the operational equation for $t_m(\va)$ of
Eq.~\eqref{190808.1} can also be rewritten as
\begin{align}
\label{201115.8}
t_m(\va)&=V-t_m(\va)[G-L_m(p,\va)]V~.
\end{align}

\section{General solution of the  PWAs  for contact interactions}
\label{sec.200406.1}
\def\theequation{\arabic{section}.\arabic{equation}}
\setcounter{equation}{0}

We consider the two-fermion scattering by a zero-range potential. We also adjust the normalization of the PWAs such it
is the same as in the ERE, that is, $\Im t(p,p)=-p|t(p,p)|^2$.
This choice is convenient since it simplifies the matching procedure with the ERE heavily used in the following. This implies
to multiply by a factor $4\pi/m$ the PWAs from the previous sections. 

\subsection{The uncoupled case}
\label{200815.1}

We start with the LS equation in partial waves,
\begin{align}
\label{200406.1a}
t(q,{r})&=v(q,{r})-v(q,q')Gt(q',{r})~,
\end{align}
where, for shortening the notation,  we do not show the integration symbol over $q'$, nor the full integrand, 
being all this understood when this continuous variable is repeated.
For instance, after including all the factors and symbols  Eq.~\eqref{200406.1a} becomes
\begin{align}
  \label{220401.1}
  t(q,r)=v(q,r)
  +\frac{2}{\pi}\int_0^\infty \frac{{q'}^2 dq' }{{q'}^2-p^2-i\ep}v(q,{q'})t({q'},r)~.
\end{align}
We now explicitly build in the momentum factors required by the centrifugal barrier potential, which  factorize in PWAs for contact interactions without left-hand cut (LC). Thus,  we write
\begin{align}
\label{200406.2c}
t(q,q')&=q^\ell {q'}^\ell \tau_\ell(q^2,{q'}^2)~,\\
v(q,q')&=q^\ell {q'}^\ell w(q^2,{q'}^2)~,\nn
\end{align}
where $w(q^2,{q'}^2)$ is a polynomial in its arguments.\footnote{The reduced potential function $w(q^2,{q'}^2)$ only depends on the square of the momenta because the full potential $q^\ell {q'}^{\ell'} w(q^2,{q'}^2)$ under the exchange $q\to -q$ and $q'\to -q'$ scales as $(-1)^{\ell}$ and $(-1)^{\ell'}$, respectively \cite{Oller:2018zts}. This is accounted for by the prefactor $q^\ell {q'}^{\ell'}$.} As a result the LS equation becomes
\begin{align}
\label{200406.1}
\tau_\ell(q^2,{r}^2)&=w(q^2,{r}^2)-w(q^2,{q'}^2){q'}^{2\ell}G\tau_\ell({q'}^2,{r}^2)~.
\end{align}

In order to calculate the on-shell $T$ matrix in the medium we also need the 
off-shell $T$ matrix
in vacuum. 
 An important remark to note in the IE for $t_m(q,p)$, Eq.~\eqref{190624.3b}, is that the off-shell momenta are bounded because the intermediate states are of an in-medium mixed type, contributing to the loop integral $L_m$ of one baryon line inside the Fermi sea, cf. Eq.~\eqref{190615.3}.
 Furthermore, once the on-shell $t_m(\vp,\vp,\va)$ is calculated the momenta involved in the Eq.~\eqref{200501.6b} for evaluating $\cE$ are bounded by the Fermi momenta because of $L_d$. Therefore, $q/\Lambda$ and $p/\Lambda$ vanish for $\Lambda\to \infty$, both in the off- and on-shell cases, respectively.  Let us notice that this is not the case in {\it vacuum}, because  when working out the off-shell  $T$ matrix from a LS equation one must consider off-shell momenta as large as the cutoff.

We first consider half-off-shell scattering, and afterwards we generalize our analysis to the off-shell case. Due to  
half-off-shell unitarity in partial waves (see e.g. Sec.~2 of Ref.~\cite{Oller:2018zts}), we can write the PWA as
\begin{align}
\label{200406.2}
\tau_\ell(q^2,p^2)&=D^{-1}(p^2)N(q^2,p^2)~,
\end{align}
Since there is no LC for a zero-range potential then $N(q^2,{q'}^2)$ is a rational function 
in its arguments, being real for real momenta  \cite{Oller:1998zr}.
Here, $D(p^2)$ is an analytic function in the cut complex $p^2$ plane with only a right-hand cut (RHC) for real and positive values of $p^2$.

For on-shell scattering $q^2={q'}^2=p^2$ and we can reabsorb $N(p^2,p^2)$ in a redefinition of $D(p^2)$, such that \cite{Oller:1998zr} 
\begin{align}
\label{200406.2b}
N(p^2,p^2)&=1~.
\end{align}
To achieve this just divide the original numerator and denominator functions in $\tau_\ell(p^2,{q'}^2)$ by $N(p^2,p^2)$. 
The possible zeros of $N(p^2,p^2)$ would
give rise to poles in the function $D(p^2)$, the Castillejo-Dalitz-Dyson poles \cite{Castillejo:1955ed,Oller:1998zr}. 

Adopting in the following the  convenient redefinition in Eq.~\eqref{200406.2b}, the imaginary part of $D(p^2)$ along the RHC becomes 
\begin{align}
\label{200406.3}
\Im D(p^2)&=-p^{2\ell}\sqrt{p^2},~p^2>0~.
\end{align}
Implementing Eq.~\eqref{200406.2} into Eq.~\eqref{200406.1},
one deduces from the latter the following IE for $N(q^2,p^2)$, 
\begin{align}
\label{200406.4}
N(q^2,{p}^2)=D(p^2)w(q^2,{p}^2)-w(q^2,{q'}^{2}){q'}^{2\ell}G N({q'}^2,p^2)~,
\end{align}
The imaginary part of this equation is zero because of Eq.~\eqref{200406.3}, taking into account
that $\Im G=-p$. This is indeed a consistency check of the general result that $N(q^2,p^2)$ has no RHC.     
Denoting the real part of $D(p^2)$ by $D_r(p^2)$,  Eq.~\eqref{200406.4} becomes
\begin{align}
\label{200406.5}
N(q^2,p^2)&=D_r(p^2)w(q^2,p^2)-w(q^2,{q'}^2){q'}^{2\ell}\Re G\, N({q'^{2}},p^2)=D_r(p^2)w(q^2,p^2)+Q(q^2,p^2)~,
\end{align}
with
\begin{align}
\label{200406.6}
Q(q^2,p^2)&=-w(q^2,{q'}^2){q'}^{2\ell}\Re G\, N({q'}^2,p^2)~,
\end{align}
which is a polynomial in the $q^2$ argument. Notice that ${q'}^{2\ell} \Re G $ is meant to represent the Cauchy principal part of the $q'$ integral involved.

To shorten the notation when considering on-shell scattering the different functions $\tau_\ell$, $Q$ and $w$ are written with only one argument, namely, as $\tau_\ell(p^2)$, $Q(p^2)$ and $w(p^2)$, respectively. 
For  on-shell scattering, in which $N(p^2)=1$, 
we can isolate $D_r(p^2)$ from Eq.~\eqref{200406.5}, which then reads
\begin{align}
\label{200406.7}
D_r(p^2)&=\frac{1-Q(p^2)}{w(p^2)}~.
\end{align}
This function is the one that is matched with the ERE,
\begin{align}
\label{201014.1}
D_r(p^2)&=-\frac{1}{a}+\frac{1}{2}rp^2+\sum_{m=2}^\infty \nu_m p^{2m}
=p^{2\ell +1} \cot \delta_\ell~.
\end{align}
Let us stress that this result is a consequence of unitarity and analyticity, with the latter exploiting the fact that no LC is present in the PWAs when considering only contact interactions.
It is entirely expressed in terms of the experimental phase shifts.

To calculate $N(q^2,{p}^2)$ we  substitute the expression for $D_r(p^2)$,
Eq.~\eqref{200406.7},  into Eq.~\eqref{200406.5}, which then reads
\begin{align}
\label{200407.8}
N(q^2,{p}^2)&=\frac{w(q^2,{p}^2)}{w(p^2)}+Q(q^2,{r}^2)-Q(p^2)\frac{w(q^2,{p}^2)}{w(p^2)}~.
\end{align}


By solving explicit examples of off-shell scattering with cutoff regularization for contact interactions with the potential $w({q'}^2,q^2)=\sum_{\alpha,\beta=0} \omega_{\alpha\beta}{q'}^{2\alpha}q^{2\beta}$ up to an including sixth degree in the arguments, we have checked that   after renormalization by matching with the ERE,\footnote{The divergent part of the integrals involved can be expressed in terms of the basic functions
\begin{align}
\label{201129.6}
I_n&=\frac{2}{\pi}\dashint_0^\Lambda\frac{q^2 dq}{q^2-p^2}q^{2n}=\int_0^\Lambda dq q^{2n}+p^2I_{n-1}~,\\
\label{201129.7}
L_n&=\frac{2}{\pi}\int_0^\Lambda dq\,q^{2n}=\theta_n\Lambda^{2n+1}~.
\end{align}
The coefficients $\theta_n$ specify the cutoff regularization scheme.}
the coefficients $\omega_{\alpha\beta}$ scale with the cutoff $\Lambda$ as
 \begin{align}
   \label{220402.1}
   \omega_{\alpha\beta}\xrightarrow[\Lambda\to\infty]{} O(\Lambda^{-2(\alpha+\beta)})~.
 \end{align}
This rule follows the dimension of $\omega_{\alpha\beta}$ corresponding to ${\rm Lenght}^{2(\alpha+\beta)+1}$, and we take it as granted in the following discussions. The previous equation also holds for a separable potential.\footnote{An explicit account of our analyses with contact-interaction potentials can be provided to the interested reader on demand.}

Importantly, Eq.~\eqref{220402.1}  allows one to conclude that
\begin{align}
\label{200407.9}
N(q^2,{p}^2)\xrightarrow[\Lambda\to\infty]{}
1~.
\end{align}
 The reasoning is the following: i) The ratio $w(q^2,{p}^2)/w(p^2)\to \omega_{00}/\omega_{00}=1$ for $\Lambda\to \infty$, and then  Eq.~\eqref{200407.8} simplifies as
\begin{align}
\label{200402.2}
N(q^2,{p}^2)&=1+Q(q^2,{p}^2)-Q(p^2)~.
\end{align}
ii) The monomial $\omega_{\alpha\beta}(q^2,{q'}^2)$ gives rise to extra cutoff powers of highest power $\Lambda^{2\beta}$ when implemented in the IE Eq.~\eqref{200406.1}, as compared with those stemming from $\omega_{00}$. However, the latter, because of its dimension ruling the scaling with the cutoff, is less suppressed precisely by ${\cal O}(\Lambda^{-2(\alpha+\beta)})$ as compared with $w_{\alpha\beta}$. Therefore, even taking into account the extra cutoff powers resulting by integrating the off-shell arguments in the IE of $\tau_\ell$, it comes out that the contributions to this IE from $w_{\alpha\beta}$ are suppressed by  ${\cal O}(\Lambda^{-2\alpha})$ for $\alpha>0$, as compared to those from $\omega_{00}$. Therefore, the contribution with $\alpha>0$ vanish in $Q(q^2,p^2)$ for $\Lambda\to \infty$. 

 However, for $\alpha=0$ there is no such suppression of the contributions arising from the integration of $w_{0\beta}(q^2,{q'}^2)$ to the left of the symbol $\Re G$ in Eq.~\eqref{200406.6}. In the same manner, when the potential $w({q'}^2,p^2)$ acts to the right in Eq.~\eqref{200406.6} (think of an iterative solution of this IE), the only  surviving contributions in the limit $\Lambda\to\infty$ correspond to $\omega_{\beta 0}$. Importantly, all these contributions involving $\omega_{0\alpha}$ to the left or $\omega_{\beta 0}$ to the right of $\Re G$ in Eq.~\eqref{200406.6} are independent of $q$, because the integration in the ${q'}$ variable is then only a function of $p^2$. 
 Thus, they cancel in the difference $Q(q^2,p^2)-Q(p^2)$ present in Eq.~\eqref{200402.2}, and
 Eq.~\eqref{200407.9} follows.

 For the off-shell case ones writes $\tau_\ell(q^2,r^2)=D^{-1}(p^2)N(q^2,r^2)$, with $r$ another momentum,  substitutes it in Eq.~\eqref{200406.1} and, by taking into account the scaling rule of Eq.~\eqref{220402.1}, one has that
 \begin{align}
N(q^2,{r}^2)=D(p^2)\omega_{00}-w(q^2,{q'}^{2}){q'}^{2\ell}G N({q'}^2,{r}^2)~.
 \end{align}
This equation has no imaginary part because of Eqs.~\eqref{200406.3} and \eqref{200407.9} (due to time-reversal invariance $N(p^2,q^2)=N(q^2,p^2)$), and then $N(q^2,r^2)$ has no RHC. Therefore, $N(q^2,r^2)$ satisfies an analogous equation to Eq.~\eqref{200402.2},
\begin{align}
  N(q^2,{r}^2)&=1+Q(q^2,{r}^2)-Q(p^2)~.
  \end{align}
Following then the same reasoning as used below Eq.~\eqref{200402.2},  one concludes that
\begin{align}
  \label{221207.1}
N(q^2,{q'}^2)\xrightarrow[\Lambda\to\infty]{}
1~.
\end{align}
 As a consequence of Eqs.~\eqref{200406.2}, \eqref{201014.1}  and \eqref{221207.1},  the vacuum 
off-shell PWAs   can be expressed directly 
in terms of the experimental phase shifts for $q,~{q'}\leq 2{\rm max}(\xi_1,\xi_2)$ as
\begin{align}
\label{201017.3}
t(q,{q'})&=\frac{(q{q'})^\ell }{p^{2\ell+1}\cot\delta_\ell-ip^{2\ell+1}}~. 
\end{align}


\subsection{The coupled case}
\label{sec.220413.1}

This section is a generalization to coupled PWAs of the results in Sec.~\ref{200815.1}, and we follow similar steps as in the uncoupled case. 
We use matrix notation which makes more straightforward this generalization process.  For $M$ coupled PWAs  the LS equation in matrix notation is written as
\begin{align}
\label{200407.16}
t(q,r)&=v(q,r)-v(q,q')Gt(q',r)~,
\end{align}
so that now $t$, and $v$ are $M\times M$ matrices and $G$ is a diagonal matrix of the same order.   
We also introduce the matrix $(q)^\ell$ which is a diagonal matrix whose $i_{\rm th}$ entry is
$q^{\ell_i}$,  being $\ell_i$ the orbital angular momentum of the $i_{\rm th}$ PWA.
The right threshold behaviors of $t(q,{q'})$ and $v(q,{q'})$ are 
explicitly taken into account analogously to Eq.~\eqref{200406.2c} by writing, respectively, 
\begin{align}
\label{200407.17}
t(q,{q'})&=(q)^\ell\tau_\ell(q^2,{q'}^2)(q')^\ell~,\\
v(q,{q'})&=(q)^\ell w(q^2,{q'}^2)(q')^\ell~.\nn
\end{align}
Multiplying Eq.~\eqref{200407.16}
by $(q)^{-\ell}$ and $({q'})^{-\ell}$ to the left and right, respectively, we have
\begin{align}
\label{200407.18}
\tau_\ell(q^2,r^2)=w(q^2,r^2)-w(q^2,{q'}^2)(q')^{2\ell} G \tau_\ell({q'}^2,r^2)~.
\end{align}
Invoking the $N/D$ method in coupled channels \cite{Bjorken:1960zz} we write
\begin{align}
\label{200407.19}
\tau_\ell(q^2,{q'}^2)&=N(q^2,{q'}^2)D^{-1}(p^2)~,\\
\Im D(p^2)&=-(p)^{2\ell+1}N(p^2)~,\nn
\end{align}
where $D(p^2)$ and $N(q^2,{q}^2)$ are $M\times M$  matrices. 
 The former has only RHC and the latter has none in the case of contact interactions. 
Both matrices of functions can be chosen
such that for on-shell scattering \cite{Oller:1998zr}
\begin{align}
\label{200407.20}
N(p^2)&=\mathbb{I}~.
\end{align}
The equation for $N(q^2,r^2)$ that results from Eq.~\eqref{200407.18} is
\begin{align}
\label{200407.21}
N( q^2 , r^2) & = w ( q^2 , r^2 ) D( p^2 ) - w ( q^2,{q'}^2 ) (q')^{2\ell} G  N({q'}^2,r^2) ~.
\end{align}
Particularizing this equation to on-shell scattering we have that
\begin{align}
\label{200407.23}
D_r(p^2)&=w(p^2)^{-1}(\mathbb{I}-Q(p^2))~,
\end{align}
where $D_r(p^2)=\Re D(p^2)$ and
\begin{align}
\label{200407.22b}
Q(q^2,r^2)&=-w(q^2,{q'}^2)(q')^{2\ell}\Re G N({q'}^2,r^2)~.
\end{align}

Equation \eqref{200407.23} is matched with the ERE in coupled channels to all orders, so that 
\begin{align}
\label{200407.24}
D_r(p^2)&=w(p^2)^{-1}(\mathbb{I}-Q(p^2))=-(a)^{-1}+\frac{1}{2}(r)p^2+\sum_{m=2}^\infty (\nu_m)p^{2m}~.
\end{align}
Here all the shape parameters are actually matrices \cite{Perez:2014waa}, a fact indicated by placing them between brackets.

Explicitly, if $S(p^2)$ is the $S$-matrix projected in partial waves, the corresponding coupled PWAs can be written as
\begin{align}
\label{201017.4}
t(p^2)&=(p)^{\ell}D^{-1}(p^2)(p)^\ell=\frac{1}{2ip}(S(p^2)-\mathbb{I})~.
\end{align}
For   vacuum $NN$ scattering,
the $S$ matrix for coupled PWAs can be expressed as 
\begin{align}
\label{201017.4b}
S&=\left(\begin{array}{ll} \cos 2\ep \,e^{2\delta_1} & i\sin2\ep \,e^{i(\delta_1+\delta_2)} \\
i\sin2\ep\, e^{i(\delta_1+\delta_2)} & \cos2\ep \,e^{2i\delta_2}
\end{array}\right)~,
\end{align}
where $\ep$ is the mixing angle and $\delta_{11}$, $\delta_{22}$ are the phase shifts for waves 1 and 2, respectively.

A completely analogous analysis as for the single-channel case implies that
$N(q^2,{q'}^2)\to \mathbb{I}$ for $\Lambda\to\infty$ (we recall that we are interested in off-shell momenta bounded by twice the largest Fermi momentum).  
Therefore, when the cutoff is sent to infinity, the expression for the off-shell coupled PWAs is
\begin{align}
\label{200407.28}
t(q,q')&=\left(\frac{q}{p}\right)^{\ell}\frac{1}{2ip}(S(p^2)-\mathbb{I})\left(\frac{q'}{p}\right)^{\ell}~.
\end{align}
In the previous expression the matrix $(q/p)^\ell$ is $(q)^\ell(p)^{-\ell}=(p)^{-\ell}(q)^\ell$.
As in the uncoupled case, Eq.~\eqref{200407.28} has been checked for  polynomial
potentials in coupled channels up to sixth degree in its arguments.

\subsection{The in-medium $T$ matrix}
\label{sec.200408.1}
For the in-medium $T$ matrix the results for the vacuum $T$ matrix expressed in
Eqs.~\eqref{201017.3} and \eqref{200407.28}  allow us to derive  an algebraic equation to
determine $t_m(q,p)$. 
 We explicitly take care of the threshold behavior for interactions without LC, and write
\begin{align}
\label{200408.1}
t_m(q,q')&=(q)^\ell \tau_m(p^2) (q')^\ell~,
\end{align}
Of course, for coupled PWAs a matrix notation analogous to that developed in Sec.~\ref{sec.220413.1} should be understood. 

Next, by taking into account the IE satisfied by $t_m$, Eq.~\eqref{190624.2}, 
it follows  that $\tau_{m}$ obeys the {\it algebraic}  matrix equation,
\begin{align}
\label{200407.30}
\tau_m(p^2)&=\tau_\ell(p^2)+\tau_m(p^2){q'}^\ell L_m {q'}^{\ell'}\tau_\ell(p^2)~,
\end{align}
with the matrix elements of  $ {q'}^\ell L_m {q'}^{\ell}$ given by 
\begin{align}
  \label{220413.1}
         [{q'}^\ell L_{m;J_2\mu_2 \ell_2,J_1\mu_1 \ell_1} {q'}^{\ell}]_{J_2\ell_2,J_1\ell_1}
=\delta_{\mu_2\mu_1}\frac{m}{(2\pi)^2}\int_0^\infty \frac{k^2dk}{k^2-p^2-i\ep}
  k^{\ell_1+\ell_2}{{\cal B}}_{J_2\mu_1\ell_2I_2,J_1\mu_1\ell_1I_1}~.
\end{align}
The matrix $[{{\cal B}}]$ is given by the matrix $[{\cal A}]$, defined in Eq.~\eqref{201114.16},  but removing the one in the factor between square brackets. Namely,
\begin{align}
  \label{220403.1}
{\cal B}_{J_2\mu\ell_2 I_2, J_1\mu\ell_1 I_1}
&=-\chi(S\ell_2 I_2)\chi(S\ell_1I_1)\sum_{m_3 s_3}
(m_3s_3\mu|\ell_2SJ_2)(m_3s_3\mu|\ell_1SJ_1) \int d\hvk Y_{\ell_2}^{m_3}(\hvk)^*
Y_{\ell_1}^{m_3}(\hvk)\nn \\
&\times \Big[\theta(\xi_1-|\vk+a\hvz|)+\theta(\xi_2-|\vk-a\hvz|)\Big]~.
\end{align}

Finally, the solution of Eq.~\eqref{200407.30} is 
\begin{align}
\label{200407.31}
\tau_m(p^2)&=\left[\tau_\ell(p^2)^{-1}- {q'}^\ell L_m {q'}^{\ell}\right]^{-1}=\left[(p)^\ell \left(\frac{2\pi(S_\ell-\mathbb{I})}{im p}\right)^{-1}(p)^\ell- {q'}^\ell L_m {q'}^{\ell}\right]^{-1}~.
\end{align}
\subsection{Uniqueness of the on-shell in-medium  $T$ matrix}
\label{sec.201018.1}

We discuss here that the  {\it on-shell in-medium PWA} $t_m(p,a\hvz)$ shown in Eqs.~\eqref{200408.1} and \eqref{200407.31} is independent of the regulator, as well as unique. 
The point is to consider the scattering amplitude of two on-shell fermions in the medium, whose imaginary part is not only due to $G$ but also to $L_m$. The former comprises the contributions from intermediate states of two fermions in vacuum, and the latter from the mixed intermediate states with one Fermi-sea insertion. 
As a result, $\tau_m(p^2)^{-1}-i(p)^\ell\Im(G(p)-L_m)(p)^\ell$ is amenable to a power expansion in $p^2$ around threshold because the branch-point singularity at $p=0$ has been removed. This is then the in-medium equivalent of the ERE in vacuum, cf. Eq.~\eqref{200407.24}. 

From this point of view, this in-medium ERE can be seen as a dressing or flow of the ERE parameters in vacuum  because of the finite density of fermions, so that one has $a(\xi_1,\xi_2)$, $r(\xi_1,\xi_2)$, and $\nu_m(\xi_1,\xi_2)$ for $m\geq 2$. In the limit $\xi_i\to 0$, of course, one has  the boundary conditions $a(0,0)=a$, $r(0,0)=r$ and, in general, $\nu_m(0,0)=\nu_m$, $m\geq 2$, where the vacuum values have been denoted with the usual  symbols.

By using cutoff regularization we have been able to work out the dressing of the ERE parameters as a function of the Fermi momenta $\xi_i$.
This is accomplished because  $t_m(p,a\hvz)$ has been calculated to all orders in the ERE, being expressed directly in terms of the vacuum phase shifts and mixing angles. 
Since any other regularization method
respecting analyticity and unitarity in the limit of contact interactions  should agree on the on-shell  $t_m(p,a\hvz)$ when taking into account all higher orders in the ERE, then it follows our claim. 

Making use of these results one can then resolve the regulator dependence \cite{Schafer:2005kg} already observed of in-medium non-perturbative calculations of $\bcE$ by performing them either with cutoff or dimensional regularization. The point here is not to perform partial calculations up to some order in the expansion of the contact interactions but to include all orders, so that the physical results are directly expressed in terms of the phase shifts and mixing angles as determined in vacuum scattering experiments. 

\section{Results}
\def\theequation{\arabic{section}.\arabic{equation}}
\setcounter{equation}{0}
\label{sec.221013.1}

We have resummed the ladder diagrams for calculating $\bcE$, Eqs.~\eqref{190615.1b} and \eqref{200501.6b}.
We have also been able to solve $t_m$ in the nuclear medium for contact interactions, cf. Eqs.~\eqref{200408.1} and \eqref{200407.31}. 
As a consequence, our results are renormalized and expressed directly in terms of the experimental phase shifts and mixing angles of $NN$ scattering \cite{Perez:2014waa}.

The calculations are based on assuming contact interactions between the two interacting nucleons, whose range of validity is limited by the onset of the left-hand cut in PWAs  due to pion exchanges, which occurs for $p^2<-m_\pi^2/4$ (this limit is determined by  one-pion exchange). Nonetheless, the on-shell PWAs in vacuum have been directly expressed in terms of the phase shifts, which is valid for all momenta. But the off-shell vacuum PWAs $t_{ij}(q,r)$, needed for in-medium calculations, are proportional to ${q}^{\ell_i}{r}^{\ell_j}$ in the  off-shell momenta, a functional form stemming from the contact-interaction nature assumed for the potential. 
Indeed, one would  expect this functional form to be valid 
only for small momentum compared with $m_\pi$.
 As said, $-m_\pi^2/4$ settles the start of the LC in the momentum-squared complex plane, 
so that the off-shell factor $q^{\ell_i}r^{\ell_j}$ would set in only as the limiting behavior for $|q|,\,|r|\ll m_\pi$. 
Therefore, we would expect that the strong off-shell dependence proportional to $k^{\ell_1+\ell_2}$ in the calculation of $L_m$  in Eq.~\eqref{220413.1} would be tamed for momenta of ${\cal O}(m_\pi)$. For instance, this is the case if one calculates the $NN$ PWAs at tree level from one-pion exchange, as given in the Appendix A of \cite{Oller:2019ssq}.

Then, we take the attitude of showing our results for low densities up to Fermi momenta $k_F= 150~\text{MeV}\sim m_\pi$, which corresponds to symmetric nuclear matter (SNM) and pure neutron matter (PNM) densities of $3\times 10^{-2}$ and $1.5 \times 10^{-2}$~fm$^{-3}$, respectively. We notice that the onset of the sensitivity to the pion LC is smooth and gradual, as it is shown by the validity of the ERE with a few terms in reproducing the $NN$  phase shifts for momenta clearly above $m_\pi/2$ \cite{Entem:2016ipb,Oller:2018zts,Guo:2013rpa,Oller:2014uxa}. Furthermore, here the use of the vacuum off-shell PWAs is always in integral expressions, so that there is an averaging process, and one is not directly sensitive to specific values of momenta.  Then, we consider reasonable to extrapolate in $k_F$  and show the results for Fermi momenta up to around $m_\pi$. Indeed, it is not uncommon for pionless EFT to show results for momenta up to around $m_\pi$ \cite{Koenig:2011lmm,Lee:2004qd,Schafer:2022hzo}.

To estimate the uncertainty in this extrapolation, we also multiply the off-shell dependence on $k$ for $k>m_\pi/2$ in the calculation of $L_m$, Eq.~\eqref{220413.1}, by the Gaussian regulator $\exp\left(-(k-m_\pi/2)^2/Q^2\right)$. The scale $Q> m_\pi$, and in this way higher values of $k$ compared with $ m_\pi/2$ are suppressed in the calculation of $L_m$  (the onset of LC in the complex $k$-plane occurs at $\pm i m_\pi/2$). Notice that this procedure is implemented only for estimating uncertainties, and our benchmark values correspond to $Q\to\infty$. The extent of the uncertainty is determined by taking the lowest value $Q=m_\pi$. 

We also consider the impact in our results of taking an effective nucleon mass $m(\rho)$  in the nuclear medium. At the level of the kinetic energy density  $\cE_K$ we have to replace $m$ by $m(\rho)$ in Eq.~\eqref{220407.3}.  For the interacting energy density, $\cE_{\cL}$, the substitution  $m\to m(\rho)$ has to be made in $L_m$ and $L_d$, but not in the calculation of the vacuum scattering PWA and, therefore, not in $(p)^\ell \left(\frac{2\pi(S_\ell-\mathbb{I})}{im p}\right)^{-1}(p)^\ell$ for $\tau_m(p^2)$, Eq.~\eqref{200407.31}. The replacement $m\to m(\rho)$ is also needed in the prefactor $-2i/(m\pi^3)$  in Eq.~\eqref{200501.6b}, since this is linked to the calculation of $L_d$. At nuclear matter saturation density $\rho_0$ Ref.~\cite{Holt:2016pjb} gives $m^*\equiv m(\rho_0)\approx 0.7~m$, while for pure neutron matter Ref.~\cite{Huth:2020ozf} obtains $m^*\approx 0.9~m$. We take these values and use for each case an extrapolation in density of the form $m(\rho)=\frac{m}{1+\frac{\rho}{\rho_0}(\frac{m}{m^*}-1)}$\, 
which becomes linear in $\rho$  for low densities. This behavior is in agreement with Refs.~\cite{Holt:2016pjb,Huth:2020ozf}, and it is also the expected leading one by having into account the self-energy corrections from $NN$ interactions \cite{Lacour:2009ej,Oller:2009zt,Meissner:2001gz}. Nonetheless, this source of estimated uncertainty is typically much smaller than the one stemming from the variation of the  Gaussian cutoff $Q$, and it is not really relevant in the presentation of our results. 


\subsection{Symmetric nuclear matter}
\label{sec.220403.2}

\begin{figure}[H]
  \begin{center}
 \includegraphics[width=0.4\textwidth]{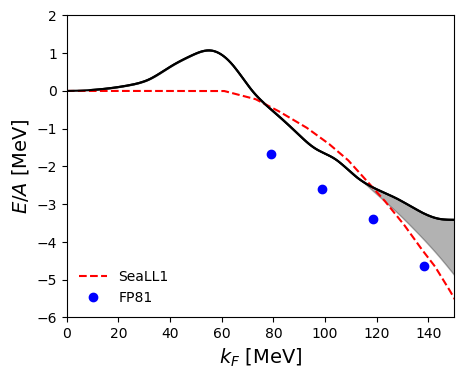}
  \caption{{\small 
  \label{fig.220823.1} The energy per nucleon for SNM as a function of $k_F<150$~MeV from the ladder resummation is given by the black solid line, and the gray area represents the estimated uncertainty. In addition, we also show the results from the variational calculation of Ref.~\cite{Friedman:1981qw} (blue filled circles), and the density functional SeaLL1  \cite{Bulgac:2017bho} (red dashed line). }}
\end{center}  \end{figure}

We show in Fig.~\ref{fig.220823.1} the resulting $\bcE$ from the ladder resummation as a function of $k_F$ by the solid black line, and the estimated uncertainty corresponds to the gray area. In the same figure we also show other low-density determinations by the blue filled circles, corresponding to the variational calculation of Ref.~\cite{Friedman:1981qw}, and the red dashed line is the result from the density functional SeaLL1 \cite{Bulgac:2017bho}. 

It is notorious that $\bcE>0$ up to $k_F\simeq 70$~MeV, or $\rho\simeq 3\times 10^{-3}$~fm$^{-3}$. This clearly indicates that SNM is not stable at such low values of the density, where the resummation of the ladder diagrams provides robust results. Of course, this phenomenon should not come as a surprise since it is well-known \cite{Shen:1998gq,Shen:1998by,Horowitz:2005nd} that at low densities the stable phase is not longer homogeneous, as $\alpha$ particles \cite{Horowitz:2005nd,Shen:1998by} and heavy nuclei \cite{Shen:1998gq,Shen:1998by} form.

\begin{figure}[H]
  \begin{center}
    \begin{tabular}{ll}
 \includegraphics[width=0.4\textwidth]{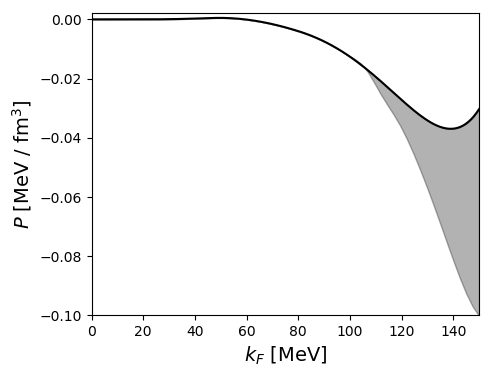} &
 \includegraphics[width=0.4\textwidth]{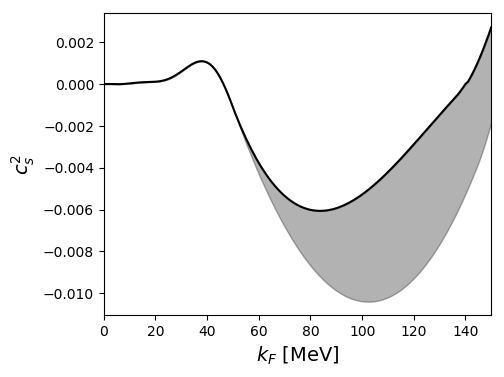}
\end{tabular}
    \caption{{\small        The pressure (left panel) and sound velocity squared (right panel) for SNM calculated from the ladder resummation are plotted as functions of $k_F<150$~MeV by the solid lines, with the estimated uncertainty given by the gray bands. \label{fig.220825.1} }}
\end{center}  \end{figure}

We can further study this region of instability of SNM, also called  spinodial region \cite{Li:1992zza}, by considering the resulting pressure, $P$, and the sound velocity squared, $c_s^2$, which respectively obey the expressions,  
\begin{align}
\label{220823.1}
P(\rho)&=\rho^2\frac{\partial\bcE}{\partial\rho}~,\\
\label{220823.2}
c_s^2(\rho)&=\frac{1}{m}\frac{\partial P}{\partial\rho}=\frac{2\rho}{m}\frac{\partial\bcE}{\partial\rho}+\frac{\rho^2}{m}\frac{\partial^2\bcE}{\partial\rho^2}~.
\end{align}

\begin{figure}[H]
  \begin{center}
    \includegraphics[width=.4\textwidth]{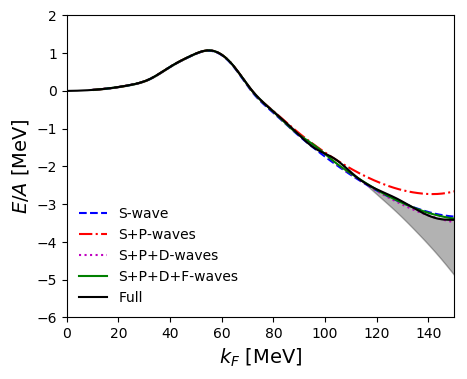}
    \caption{{\small
        The partial-wave contributions to $\bcE$ for SNM with different orbital angular momenta are added separately. Contributions with the $S-$, $P-$, $D-$, and $F-$waves consecutively added correspond to the blue dashed, red dashed-dotted, magenta dotted, and green solid lines, respectively.  The full result is the black solid line. See the text for further details. } \label{fig.220829.1}
    }    
    \end{center}
  \end{figure}

$P(\rho)$ and  $c_s^2(\rho)$ are shown in the left panel and right panels  of Fig.~\ref{fig.220825.1}, respectively. The pressure is positive up to $k_F=59$~MeV, corresponding to a system which tends to split apart. $c_s^2$ becomes positive in the region of negative $P(\rho)$ only above a critical value of $k_F$, which we denote as $\xi_c$, and for which the resummation of ladder diagrams yields $\xi_c =139$~MeV, with around a $+20$~MeV of uncertainty. Let us recall the relation between the compressibility coefficient $K$ and $c_s^2$,
\begin{align}
\label{220823.3}
K=\frac{c_s^2}{m \rho}~,
\end{align}
so that when $c_s^2<0$ then $K<0$.  The Fermi momentum $\xi_c$ is the critical density above which the system leaves the instability region, and SNM becomes a homogeneous stable phase.
Reference~\cite{Shen:1998gq} employing a relativistic mean field theory obtained a value for the critical density around $10^{14}$~g/cm$^3$, corresponding to $\xi_c=190$~MeV. The boson-exchange model for nuclear interactions used to apply the Dirac-Bruckner approach to calculate $\bcE$ in Ref.~ \cite{Li:1992zza} gives $\xi_c\approx 200$~MeV.

We separately show in Fig.~\ref{fig.220829.1} the contributions from PWAs involving different orbital angular momenta, and their sum up to the final result given by the black solid line.  We organize the different contributions according to the mixing of PWAs in vacuum. In this way, the blue dashed line is dubbed to correspond to the $S$-waves but, because of the mixing between the ${^3S}_1$ and ${^3D}_1$ PWAs, we are actually keeping the ${^1S}_0$, and ${^3S}_1-^3D_1$ PWAs. Let us recall that we directly take the experimental phase shifts and mixing angles, so that the mixed PWAs in vacuum must be kept together. Similarly, by the notation of $P$ waves (red dash-dotted line) we are adding the PWAs ${^1P}_1$, ${^3P}_0$, ${^3P}_1$, ${^3P}_2-{^3F}_2$. For the name $D$ waves (magenta dotted line) we have in addition the ${^1D}_2$, ${^3D}_2$, ${^3D}_3-{^3G}_3$, and for the $F$ waves (green solid line) we have added the contributions from the ${^1F}_3$, ${^3F}_3$ and ${^3F}_4-{^3H}_4$ PWAs. 
 As expected, we see from Fig.~\ref{fig.220829.1} that the main contributions arise from the $S$ waves, but the $P$ waves gives a noticeable repulsion, which is compensated to large extend by the $D$-wave contributions. The convergence is already achieved with the $F$-wave contributions, being almost indistinguishable with the final curve including the $G$ waves.

\subsection{Pure neutron matter}
\label{sec.220403.1}

\begin{figure}[H]
  \begin{center}
    \includegraphics[width=.4\textwidth]{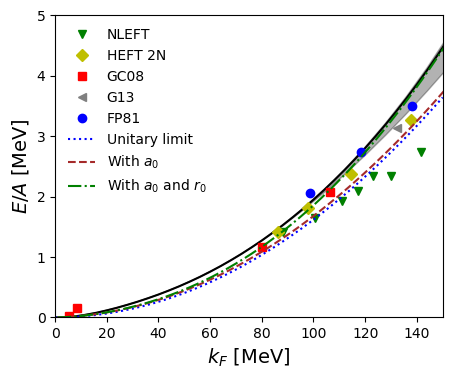}
    \caption{{\small    
$\bcE$ for PNM is plotted as a function of $k_F<150$~MeV. Our results are given by the black solid line, together with the gray band giving the uncertainty estimated. The resulting curve in the unitary limit $a_0=\infty$ is  the blue dotted line. We also provide the results calculated with the $^1S_0$ scattering length only, brown dashed line, plus the effective range, green dash-dotted line \cite{Alarcon:2021kpx}.  For comparison we show results for NLEFT \cite{Epelbaum:2009rkz} (green downwards triangles), HEFT 2N \cite{Wlazlowski:2014jna} (yellowish green diamonds), the two quantum Monte-Carlo calculations of Gezerlis \& Carlson \cite{Gezerlis:2007fs} (red squares) and Gezerlis \cite{Gezerlis:2013ipa} (gray left-pointing triangles), as well as the variational one of Ref.~\cite{Friedman:1981qw} (blue circles).   
    \label{fig.220824.pnm}}}
\end{center}  
  \end{figure}

\begin{figure}[ht]
  \begin{center}
    \begin{tabular}{ll}
 \includegraphics[width=0.4\textwidth]{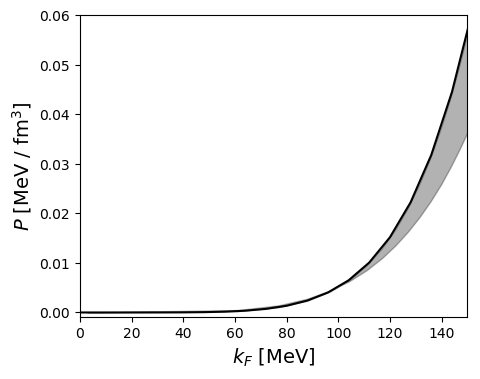} &
 \includegraphics[width=0.4\textwidth]{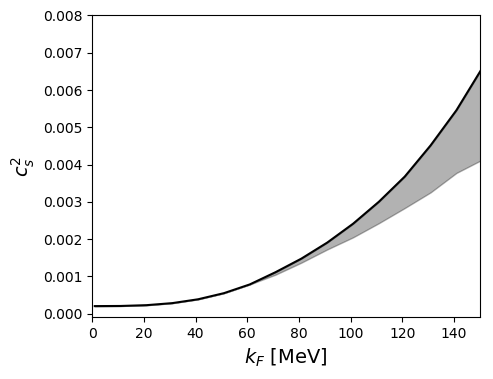}
\end{tabular}
    \caption{{\small  The pressure $P(\rho)$ (left panel) and sound velocity squared $c_s^2(\rho)$ (right panel) that result from the ladder resummation for PNM are plotted by the solid lines.
        \label{fig.220825.2} }}
\end{center}  \end{figure}

The results for $\bcE$  of PNM by resumming the ladder diagrams are shown in Fig.~\ref{fig.220824.pnm} by the solid line, with an estimated uncertainty given by the gray band.
By considering only the $S$-wave contributions, namely the PWA $^1S_0$, we plot $\bcE$  in the unitary limit (infinite scattering length) by the blue dotted line. When taking the actual value for the scattering length of the PWA $^1S_0$, $a_0=-18.95~$fm, the brown dashed line results and, after  the effective-range contributions are added with $r_0=2.75~$fm, we have the green dot-dashed line. 
The last two cases were already calculated by us in Ref.~\cite{Alarcon:2021kpx}. 
In addition we also compare with other calculations.
 The green downwards  triangles give the low-density results from nuclear lattice EFT of Ref.~\cite{Epelbaum:2009rkz}, and the blue filled circles correspond to the  variational calculation of Ref.~\cite{Friedman:1981qw}. We also show the quantum Monte Carlo results of Refs.~\cite{Gezerlis:2007fs} (red squares), and \cite{Gezerlis:2013ipa} (gray left-pointing triangles), and the auxiliary-field quantum Monte Carlo calculation of \cite{Wlazlowski:2014jna} (light green diamonds).  We see that $\bcE$ for PNM obtained from the resummation of the ladder diagrams is more repulsive than any of the other calculations shown for $k_F\gtrsim 120$~MeV. 

 We also plot the pressure $P$ and the sound velocity squared $c_s^2$ for PNM that result from the resummation of the ladder diagrams in the left and right panels of Fig.~\ref{fig.220825.2}, respectively, with the gray bands giving the estimated uncertainty as discussed above.

\begin{figure}[ht] 
  \begin{center}
\includegraphics[width=.4\textwidth]{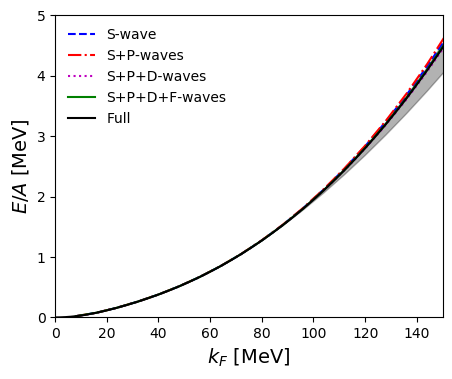}
\caption{{\small 
    The partial-wave contributions to $\bcE$ for PNM with different orbital angular momenta are added separately. The notation is the same as employed in Fig.~\ref{fig.220829.1} for SNM. 
 } \label{fig.220829.2}
}    
    \end{center}
\end{figure}

Separated partial-wave  contributions to $\bcE$ for PNM are shown in Fig.~\ref{fig.220829.2}, similarly as in Fig.~\ref{fig.220829.1} for the case of SNM. Then, by $S$-waves  (blue dashed line) we mean the contributions from  the ${^1S}_0$ PWA; $P$-waves (red dotted line)  comprise in addition those from the ${^3P}_0$, ${^3P}_1$ and ${^3P}_2-{^3F}_2$; $D$-waves (magenta dash-dotted) include the ${^1D}_2$; and $F$-waves (green solid line) comprise the contributions from the ${^3F}_3$ and ${^3F}_4-{^3H}_4$ PWAs. In the interval of values of $k_F$ shown it is clear that the full result (black solid line) is overwhelmingly dominated by the ${^1S}_0$ PWA, with a small repulsive $P$-wave contribution, which is compensated by the $D$ and higher partial waves. The convergence with the full result is reached  with almost indistinguishable $F$-wave contributions.

\begin{figure}[ht]
\begin{center}
  \includegraphics[width=.4\textwidth]{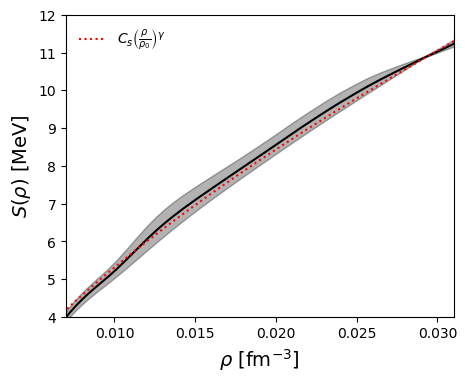}
  \caption{{\small 
     The solid line and its uncertainty band correspond to the symmetry energy $S(\rho)$ obtained from the resummation of the ladder diagrams as a function of density $\rho$. The red dashed line is the parameterization of Eq.~\eqref{220824.1} using the fitted central values of $C_s$ and $\gamma$ in Eq.~\eqref{221012.1}.}
  \label{fig.220830.1}
  }
\end{center}
\end{figure}

 By assuming a quadratic dependence of $\bcE$ on the proton fraction $x_p\equiv \rho_p/\rho$ (width $\rho_p$ the density of protons), 
we can calculate from our results the symmetry energy $S(\rho)$ as \cite{Grasso:2018pen} 
\begin{align}
  \label{220411.1}
  S(\rho)&=\bcE(\rho,0)-\bcE(\rho,\frac{1}{2})~.
\end{align}
Here, we follow the notation $\bcE(\rho,x_p)$ for $\bcE$ as a function of density and proton fraction, so that  $\bcE(\rho,\frac{1}{2})$ corresponds to SNM, and $\bcE(\rho,0)$ does for PNM.  Note that, in order to apply  Eq.~\eqref{220411.1}, one is  taking the difference of energies per nucleon at a fixed value of density $\rho$. Then, $k_F$ for SNM is a factor $1/2^{1/3}$ smaller than $k_F$ for PNM when calculating the difference $S(\rho)=\bcE(\rho,0)-\bcE(\rho,0)$. From our results shown in Figs.~\ref{fig.220824.pnm} and \ref{fig.220823.1} for PNM and SNM, respectively, we calculate $S(\rho)$, which is plotted in Fig.~\ref{fig.220830.1} by the solid line, with the gray area giving the uncertainty estimated. 

 Going on with the  quadratic dependence on the proton fraction for $\bcE(\rho,x_p)$, Ref.~\cite{Gandolfi:2009nq} wrote also the parameterization
 \begin{align}
   \label{220824.1}
  \bcE(\rho,x_p)=\bcE(\rho,\frac{1}{2})+C_s\left(\frac{\rho}{\rho_0}\right)^{\gamma_s}(1-2x_p)^2~. 
\end{align}
 In using this formula we take  $\rho_0=0.16$~fm$^{-3}$, the standard value for nuclear matter saturation. 

By fitting $S(\rho)$, shown in Fig.~\ref{fig.220830.1}, with Eq.~\eqref{220824.1} within the density range $\rho\in[0.7,3.1]\times 10^{-2}$~fm$^{-3}$ the free parameters $C_s$ and $\gamma$ are then determined. In the chosen density range $S(\rho)$ has a smooth behavior, once the region around the maximum of $\bcE$ for SNM in Fig.~\ref{fig.220829.1} is clearly left behind. 
 The values obtained from the fit are 
 \begin{align}
   \label{221012.1}
   C_s&=34.77\pm 0.15~\text{MeV}~,\\
   \gamma&=0.667\pm 0.003.\nn
 \end{align}
 By employing the central values and the parameterization of Eq.~\eqref{220824.1} we obtain the red dashed line plotted in Fig.~\ref{fig.220830.1}, which shows that the fit closely reproduce the results from the ladder resummation for $\rho\in[0.7,3.1]\times 10^{-2}$~fm$^{-3}$. 

\begin{figure}[ht]
  \begin{center}
    \begin{tabular}{ccc}
  \includegraphics[width=.5\textwidth]{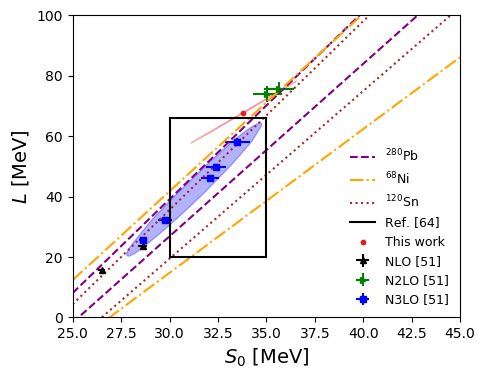}
  \end{tabular}
\end{center}
  \caption{{\small 
      $L$ versus  $S_0$   at $\rho_0$. We show our results by the red circle and its two sigma uncertainty area. We also give the outcome of Ref.~\cite{Holt:2016pjb} at different orders in the perturbative chiral expansion. The two sigma uncertainty area for the N3LO result is shown in blue. Three empirical bands, obtained in Ref.~\cite{Roca-Maza:2015eza} by analyzing the static dipole polarizability in the nuclei {$^{208}$Pb}, {$^{68}$Ni}, and {$^{120}$Sn}, are plotted. The inferred values from the same reference are also report by the square.}
\label{fig.220411.1} 
}
\end{figure}

The parameterization in Eq.~\eqref{220824.1} fixed at low densities allows us to  extend the results to larger values of $\rho$ and, in particular, consider the  values for the symmetry energy at saturation $S_0\equiv S(\rho_0)$,
\begin{align}
   \label{220824.2}
  S_0&=\bcE(\rho_0,0)-\bcE(\rho_0,\frac{1}{2})~,
\end{align}
and its slope 
    \begin{align}
      \label{220411.2}
  L&=3\rho_0\left.\frac{dS(\rho)}{d\rho}\right|_{\rho_0}=3C_s\gamma_s~.
    \end{align}
    These are magnitudes of  phenomenological interest, with a special attention devoted on the investigation of existing correlations between these quantities
    (defined and computed in infinite nuclear matter) and measured observables. Among the latter we have those in finite nuclei, such as the neutron skin thickness in neutron-rich nuclei and the electric dipole polarizability, and others in astrophysics e.g. concerning neutron stars and heavy-ion collisions with radioactive beams  \cite{Grasso:2018pen,Ghosh:2022lam,Roca-Maza:2015eza,Oertel:2016bki}.

  We show our central values for $S_0$ and $L$ in Fig.~\ref{fig.220411.1} by the red circle corresponding to $S_0=33.77$~MeV and $L=67.59$~MeV. The two-sigma uncertainty area is also given, extending over the values $31.10\leq S_0\leq 36.57$~MeV and $57.82\leq  L\leq 78.29~$MeV. The figure also gives the empirical bands obtained in Ref.~\cite{Roca-Maza:2015eza} by analyzing the data on the electric dipole polarizabilities of {$^{68}$Ni}, {$^{120}$Sn}, and {$^{208}$Pb} employing several density functionals. The same reference infers the intervals of values  $30\leq S_0\leq 35$~MeV and  $20\leq L\leq 66$~MeV, represented by the square in Fig.~\ref{fig.220411.1}, which is compatible with our calculation. In addition, we also show the outcome of Ref.~\cite{Holt:2016pjb} obtained by employing chiral perturbation theory at different orders, as indicated in the figure. The two-sigma correlation area at N3LO of Ref.~\cite{Holt:2016pjb} is given by the blue area, and lying quite close to our outcome.

\section{Conclusions}

We have studied infinite nuclear matter by resumming the series of ladder diagrams following the results of Ref.~\cite{Alarcon:2021kpx}. The master formula there given allows one to consider arbitrary nucleon-nucleon ($NN$) interactions in vacuum. This formalism can be explicitly solved for the case of $NN$ interactions driven by contact-interaction potentials. The partial-wave amplitudes up to an including $G$ waves are considered for symmetric and pure neutron matter for Fermi momentum up to 150~MeV, so that the results are convergent under the inclusion of higher partial-wave amplitudes.  The energy per particle $\bcE$ obtained from the ladder series is renormalized, without any dependence on arbitrary scales, like cutoffs or regulators, and it is directly expressed in terms of the experimental $NN$ phase shifts and mixing angles, reducing the systematic errors in the calculation of dilute nuclear matter. The knowledge of $\bcE$ as a function of density $\rho$ allows also to study other interesting observables like the pressure $P(\rho)$ (equation of state), and the sound velocity $c_s(\rho)$.

We notice that our results are specially suitable  in the low density region where a pionless description of $NN$ interactions can make sense. They comprise the  full vacuum  $NN$ interactions and the leading order nonperturbative in-medium contributions, according to the power counting of Ref.~\cite{Oller:2009zt}. An interesting application of these results given by the resummation of the ladder diagrams would be to use them as low-density constrains to the equation of state. This is specially interesting for the  calculation of the properties of neutron stars. A work on that direction is in progress.

\section*{Acknowledgements}

We would like to thank interesting discussions with Felipe J. Llanes-Estrada and Eva Lope-Oter. 
This work has been supported in part by the  MICINN AEI (Spain) Grants PID2019-106080GB-C21/AEI/10.13039/501100011033, PID2019-106080GB-C22/AEI/10.13039/501100011033, and by EU Horizon 2020 research and innovation program, STRONG-2020 project, under grant agreement No 824093.

\bibliographystyle{h-physrev5}
\bibliography{nuclearmatter}

\end{document}